# Strong Electron-Phonon Coupling and Lattice Dynamics in One-Dimensional [(CH$_3$)$_2$NH$_2$]PbI$_3$ Hybrid Perovskite

A. Nonato[1], Juan S. Rodríguez-Hernández[2], D. S. Abreu[2], C. C. S. Soares[2], Mayra A. P. Gómez[2], Alberto García-Fernández[3], María A. Señarís-Rodríguez[3], Manuel Sánchez Andújar[3], A. P. Ayala[2], and C.W.A. Paschoal[2], Rosivaldo Xavier da Silva[4,*].

[1]Coordenação de Ciências Naturais, Universidade Federal do Maranhão, Campus do Bacabal, 65700-000, Bacabal - MA, Brazil.

[2]Departamento de Física, Universidade Federal do Ceará, Campus do Pici, 65455-900, Fortaleza - CE, Brazil.

[3]University of A Coruna, UdcSolids Group, Dpt. Chemistry, Faculty of Science and Centro Interdisciplinar de Química e Bioloxía (CICA), Zapateira, 15071 A Coruña, Spain.

[4] Centro de Ciência e Tecnologia em Energia e Sustentabilidade, Universidade Federal do Recôncavo da Bahia, Feira de Santana, Bahia 44085-132, Brazil.

[*]Corresponding author. Tel: +55 98 9 98223 4219

E-mail address: rosivaldo.xs@ufrb.edu.br

# Abstract

Hybrid halide perovskites (HHPs) have attracted significant attention due to their remarkable optoelectronic properties that combine the advantages of low cost-effective fabrication methods of organic-inorganic materials. Notably, low-dimensional hybrid halide perovskites including two-dimensional (2D) layers and one-dimensional (1D) chains, are recognized for their superior stability and moisture resistance, making them highly appealing for practical applications. Particularly, $DMAPbI_3$ has attracted attention due to other interesting behaviors and properties, such as thermally induced order-disorder processes, dielectric transition, and cooperative electric ordering of DMA dipole moments. In this paper, we investigated the interplay between low-temperature SPT undergone by the low-dimensional (1D) hybrid halide perovskite-like material $DMAPbI_3$ and its optoelectronic properties. Our approach combines synchrotron X-ray powder diffraction, Raman spectroscopy, thermo-microscopy, differential scanning calorimetry (DSC), and photoluminescence (PL) techniques. Temperature-dependent Synchrotron powder diffraction and Raman Spectroscopy reveal that the modes associated with I-Pb-I and $DMA^+$ ion play a crucial role in the order-disorder SPT in $DMAPbI_3$. The reversible SPT modifies its optoelectronic properties, notably affecting its thermochromic behavior and PL emission. The origin of the PL phenomenon is associated to self-trapped excitons (STEs), which are allowed due to a strong electron-phonon coupling quantified by the Huang-Rhys factor ($S = 97 \pm 1$). Notably, we identify the longitudinal optical (LO) phonon mode at 84 $cm^{-1}$ which plays a significant role in electron-phonon interaction. Our results show these STEs not only intensify the PL spectra at lower temperatures but also induce a shift in the color emission, transforming it from a light orange-red to an intense bright strong red.

# I. INTRODUCTION

Hybrid halide perovskites (HHPs) have attracted significant attention due to their remarkable optoelectronic properties that combine the advantages of low cost-effective fabrication methods of organic-inorganic materials, [1–3] with highly efficient electrical conductivity, which makes them promising for photovoltaic technology, [4] light emitting diode (LED) applications [5] and X-ray detectors. [6–8] The promising applications of these materials are attributed to their unique optoelectronic properties, such as high optical absorption across a wide spectrum and tunable bandgaps, facilitated by the compositional versatility and dimensional variety of the organic-inorganic hybrid structure.[9–11]

To enhance the stability of HHPs, cation engineering has emerged as a strategic approach to precisely modulate the relationship between crystal structure and properties, ultimately leading to improved optoelectronic performance.[3] The crystal structure of HHPs is defined as a network of corner-sharing $BX_6$ octahedra with a general chemical formula $ABX_3$.[12] In these materials, the A site is normally occupied by a small organic alkylammonium cation or large alkali cation, B by divalent transition metal cations or a mixture of monovalent and trivalent cations and X by halogens.[3,13] Particularly, a recent focus has been the exploration of lead halide perovskites with the formula $[AmH]PbX_3$ (AmH= $[CH_3NH_3]^+$ methylammonium (MA) cation and $[CH(NH_2)_2]^+$ formamidinium (FA) cation, and X = $Cl^-$, $Br^-$ and $I^-$). This interest originates primarily from the remarkable photoconductivity observed in methylammonium lead triiodide perovskite $[CH_3NH_3]PbI_3$ ($MAPbI_3$), with a three-dimensional (3D) framework. This material has shown promise due to its solar energy conversion efficiency reaching 22.7%, surpassing other device-based materials such as quantum dots, organic and amorphous silicon solar cells, and other emerging PV technologies. Despite the promising applications of halide perovskites, several major challenges still hinder their practical implementation, such as their poor stability under ambient conditions, particularly in moist environments.[14–16]

On the other hand, notably low-dimensional inorganic-organic hybrid halide perovskites including two-dimensional (2D) layers and one-dimensional (1D) chains, are recognized for their superior stability and moisture resistance, making them highly appealing for practical applications.[17] Additionally, low-dimensional perovskites, owing to their multilayered arrangement, demonstrate distinct properties when compared to their 3D analogous. Their inherent confining electrons configuration produces stable excitons, able to interact more strongly with phonons, spins, and defects.[18] Thus, it has been shown that the broadband light emission intensity of low-dimensional perovskite is positively related to the quantum confinement effect in distorted inorganic species. Therefore, extensive attention has been also drawn to these materials due to their intrinsic broadband light emissions, primarily resulting from self-trapped excitons (STEs) induced by the significant electron-phonon coupling effect in the soft crystal lattice.[19–21] Further, STE formation is predicted to strongly correlate with system dimensionality, being more favorable in low-dimensional materials.[22]

In this context, recent studies have investigated the structural phase transition (SPT) and ionic transport properties of the new compound $DMAPbI_3$, which adopts a one-dimensional (1D) 2H-hexagonal perovskite structure. The crystal structure of the 2H-polytype consists of infinite chains of face-sharing $[PbI_6]^{4-}$ octahedra running along the c-axis, while $DMA^+$ organic cations occupy the cavities delimited by the chains of the inorganic framework, see Figure 1. The $DMAPbI_3$ undergoes SPT at $T_c = 250$ K from monoclinic symmetry with space group (SG): *P2$_1$/c* (*LT-phase*) to hexagonal symmetry with SG: *P6$_3$/mmc* at room temperature (*HT-phase*), see Figure 1. This transition is governed by two cooperative processes: (1) specific orbital interaction between the 6s orbital of $Pb^{2+}$ cation and 5p orbitals of the $I^-$ anion, allowing off-center shift of the $Pb^{2+}$ cations and (2) order-disorder process of the polar $DMA^+$.[23] Furthermore, the off-center shift of the $Pb^{2+}$ ion has been associated with the presence of stereochemically active $6s^2$ electron lone pairs, which have been linked to playing a crucial role in the lattice instability and optoelectronic properties of perovskites.[24,25] This means that the interaction of lone pairs with anions regulates energy bands, contributes to tolerance factors, enhances electronic mobility, and makes the bandgap sensitive to pressure and temperature variations.[6]

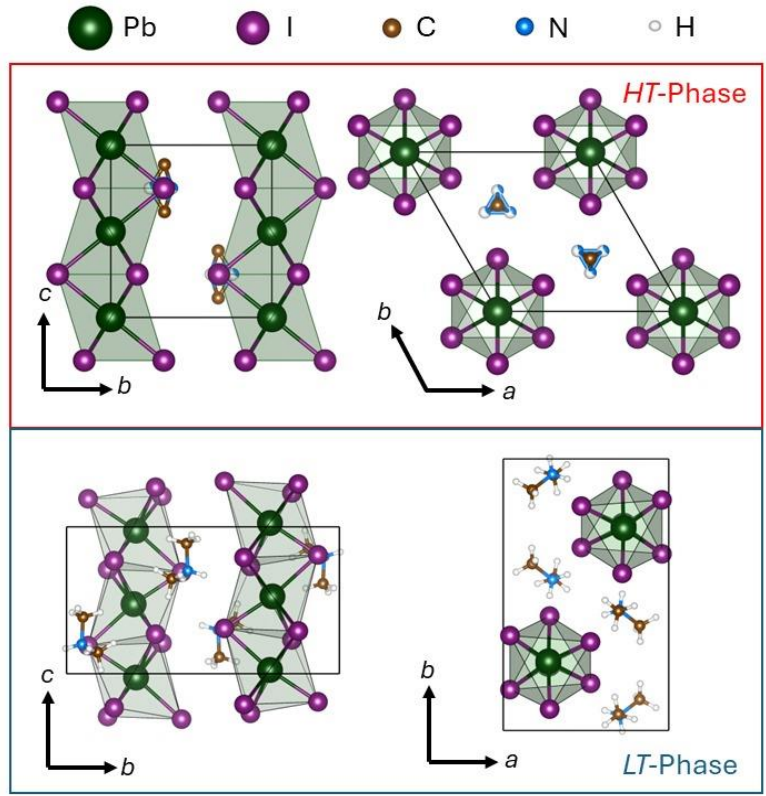

**Figure 1** – Crystalline structures of HT-phase (top) and LT-phase (bottom) for the DMAPbI$_3$ compound viewed along different directions.

Moreover, DMAPbI$_3$ has attracted attention due to other interesting behaviors and properties, such as thermally induced order-disorder processes, dielectric transition, and cooperative electric ordering of DMA dipole moments.[23] Exceptionally, due to its highly distorted [PbI$_6$]$^{4-}$ octahedra at low temperatures, DMAPbI$_3$ displays strong orbital mixing, which, combined with its low dimensionality, favors the emergence of STEs. This makes it potentially promising for uncovering new optoelectronic properties. Since the optoelectronic properties in these materials strongly depend on the SPT, comprehending their performance and stability in low-temperature conditions is imperative for their practical applications. [26,27] In this sense, Raman spectroscopy is a very sensitive and effective tool for investigating SPT, particularly in detecting lattice distortions in modes associated with PbI$_6$ octahedra, which are fundamentally linked to the electronic properties in this material. Furthermore, studying lattice dynamics enables the characterization of the temperature-dependent phonon lifetime behavior, which allows precise identification of modes associated with short- and long-range order-disorder processes that drive the SPT in DMAPbI$_3$, as well as the reorientation processes involved.

Otherwise, Fourier-transform infrared reflectivity (FTIR) measurements allow for the identification of the polar phonons that contribute most strongly to the intrinsic dielectric constant, shedding light on the frequency of the longitudinal optical phonon (e.g. LO mode) that enables strong electron-phonon coupling in DMAPbI$_3$.

Despite substantial progress in improving the efficiency of perovskite-based devices, the fundamental relationship between SPT mechanisms and the resulting optoelectronic phenomena remains a rich area for further exploration.[28] While the SPT in DMAPbI$_3$ was previously reported by Garcia *et al.*,[23] our investigation focuses on investigating its thermal evolution and implications for optoelectronic properties. We investigated how the SPT affects the electronic properties of the low-dimensional hybrid perovskite DMAPbI$_3$. Using temperature-dependent synchrotron X-ray powder diffraction (VT-SXRPD) experiments, we gain more insight into the SPT and thermomechanical response of both LT- and HT-phases. Additionally, temperature-dependent Raman spectroscopy allowed us extensively to characterize the SPT, highlighting the effects of lattice and octahedral distortions, as well as order-disorder and reorientational processes. Furthermore, through temperature-dependent photoluminescence analysis combined with FTIR measurements, we observed a strong correlation between the mechanisms in emission spectra broadening and the electron-phonon coupling effect at low temperatures. These results enhance our understanding of how optical polar phonons and lattice distortions may influence the mechanisms that enhance optoelectronic effects in these systems.

## II. EXPERIMENTAL SECTION

**Synthesis Method:** Single-crystals of both DMAPbI$_3$ were obtained using the crystallization method from a N-N-Dimethylformamide (DMF) solution. Stoichiometric amounts of PbI$_2$ (4 mmol) and DMAI (4 mmol) were dissolved by adding 5 ml of DMF. The obtained solutions were filtered through 0.45 μm PTFE filters to minimize nucleation sites. Yellow needle-shaped single crystals of DMAPbI$_3$ were obtained upon slow evaporation of the solvent at room temperature after several days. Polycrystalline powders of DMAPbI$_3$ were obtained by placing the single crystal in an agate mortar and carefully grinding it with a pestle.[29–31]

***Powder X-Ray Diffraction***: Powder X-ray diffraction analysis was performed using a Bruker D8 Advance instrument equipped with Cu-Kα radiation. The analysis spanned a range from 5° to 100°, with a step size of 0.02° and an exposure time of 5 seconds per step. The results were cross-referenced with data reported by Garcia *et al.*,[31] in the ICSD: 243860. The Rietveld refinement process was meticulously executed using the EXPO2013 software,[32] and further details can be found in **Figure S1** within the supporting information (S.I).

***Synchrotron X-Ray Powder Diffraction:*** Variable temperature synchrotron X-ray powder diffraction (SXRPD) patterns were collected at the BL04 MSPD beamline of the ALBA Synchrotron (Cerdanyola del Vallès, Spain). The wavelength and intrinsic peak shape parameters were determined by the refinement of a diffraction pattern collected from a known Si NIST standard. The X-ray wavelength used was λ= 0.412483 Å. The sample was load in a borosilicate capillary (ϕ = 0.5 mm) and rotated during data collection. The variable temperature SXRPD data were collected on heating over the temperature ranges 120–350 K using Oxford Cryostreams. Rietveld analysis was carried out using the program GSAS-II.[33]

***Infrared Reflectivity:*** The infrared reflectivity (IR) spectra were acquired with a Bruker Vertex 70V Fourier-transform spectrometer. Data in both far- and mid-infrared regions were collected using a silicon carbide (SiC) lamp (Globar) as the light source, while the signal detection was carried out using DLaTGS pyroelectric detectors. The dataset was obtained through 256 scans, and a spectral resolution of 2 cm$^{-1}$ was achieved using a wide-range silicon beam splitter.

***Optical Absorption***: In an adapted configuration, the optical absorption measurements for the DMAPbI$_3$ compound were conducted using the transmittance percentage (%T) function on the Cary 7000 spectrophotometer equipped with the Universal Measurement Accessory (UMA). The sample was positioned at an angle of 0°, while the detector was set at 180° to maximize light absorption in the compound. The spectrophotometer utilized a grating changeover at 800 nm, operated in dual-beam mode, and covered a wavelength range from 400 nm to 800 nm. The sample was a 2.5 cm diameter pellet with a thickness of 0.0124 cm.

***Differential scanning calorimetric (DSC)***: Thermal analyses were conducted in a Netzsch Maia 200 F3. The samples were heated and cooled under a nitrogen atmosphere for several cycles at 5-10 K.min$^{-1}$.

***Thermo-microscopy:*** Thermo-microscopy was conducted using an optical polarizing hot-stage microscopy station at low temperatures. The images were captured within a temperature range from 300 down to 125 K, applying a cooling rate of 5 K/min and using a 10x magnification lens. These images were recorded using a QICAM (Fast1394) camera and processed with Linkesys32 software. The temperature of the sample was controlled using the THMS600 LINKAM hot-stage station. Each photograph was taken after waiting about 4 seconds to stabilize temperature and attain an equilibrium structure. The CellProfiler [34] software was used to edit the images.

***Raman spectroscopy at low-temperatures:*** Raman spectra of DMAPbI$_3$ powder sample were collected using a high-resolution LabRam HR Horiba 800 spectrometer using a grating with 1800 lines per millimeter and equipped with a CCD detector cooled with liquid nitrogen. The 632.5 nm (red) line from the He-Ne gas laser was used as the excitation source, and the beam was focused onto the sample using a long working distance 20x/0.40 microscope objective. The experiments were carried out in the temperature range from 300 K down to 140 K. For this purpose, the samples were placed in a Nitrogen flow Linkam cryostat, whose temperature was controlled by the Linksys software with a precision of 1 K. All spectra were fitted with the Fityk software. [35]

***Photoluminescence (PL) at low-temperatures:*** The PL spectra were collected using a T64000 Jobin–Yvon spectrometer equipped with an Olympus microscope and an LN$_2$-cooled CCD to detect the emitted radiation of the sample in a single mode. The spectra were excited with an external lamp (405 nm) using a long working distance plan-achromatic objective of 20x. The temperature-dependent PL spectra were obtained by keeping the sample in a vacuum inside a He-compressed closed-cycle cryostat. A Lakeshore 330 controller controlled the temperature, keeping the precision around 0.1 K. The PL curves were adjusted with a Pearson7 fit profile using Fityk following the quantitative analysis of emission spectra by Kambhampati & Mooney [36].

## III. RESULTS

### A. Structural and thermomechanical analysis

We performed temperature-dependent synchrotron X-ray powder diffraction (VT-SXRPD) experiments to gain more insight into the phase transition and thermomechanical properties of both LT- and HT-phases. Figure 2 shows the obtained patterns, revealing a rather abrupt structural phase transition occurring at $T_c \approx 245$ K, in good agreement with the DSC results. For T > 250 K, the only phase present is the one with hexagonal symmetry (HT-phase), while for T < 240 K, the only phase present is the one with monoclinic symmetry (LT-phase). It is worth noting that there is no coexistence of both phases at any temperature range.

Additionally, the obtained patterns were fitted using Rietveld refinement (see Figure S2 of S.I), and the lattice parameters and angle obtained from these refinements are shown in Figure 2. As we can observe, the temperature dependence of the lattice parameters exhibits a notable anisotropic response. For the LT-phase, the values of the lattice parameters along the $b$- and $c$-axis increase upon heating, while the value along the $a$-axis is almost temperature-independent. Additionally, there is a remarkable decrease in the beta angle upon heating. In the case of the HT-phase, the value of the $a$-axis increases, while the value along the $c$-axis decreases upon heating.

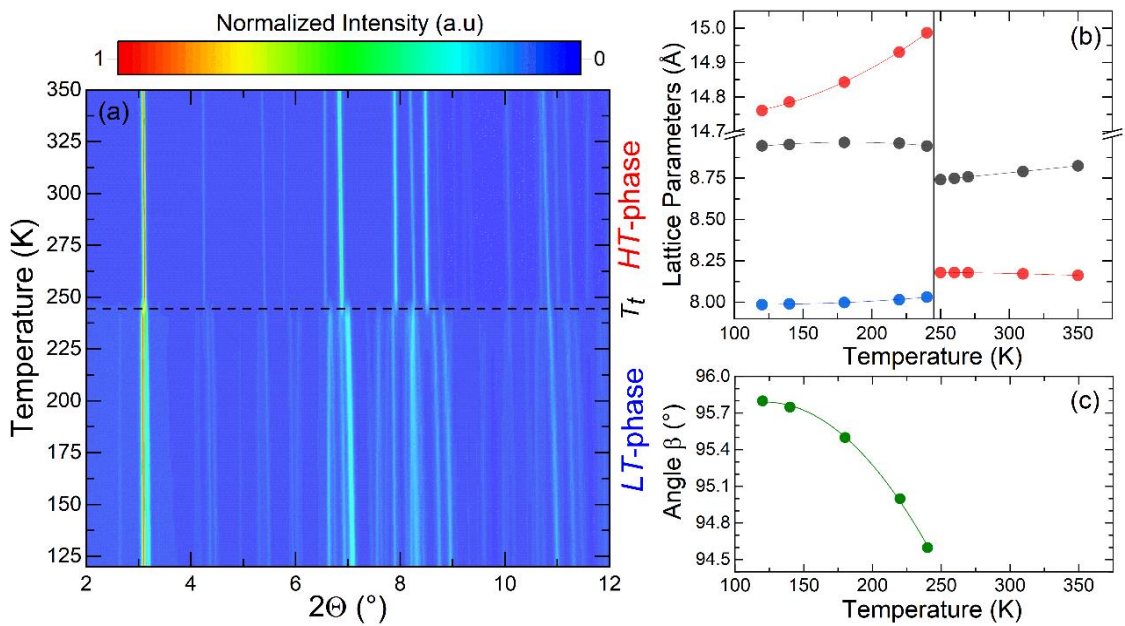

**Figure 2** – (left) Temperature-dependent synchrotron X-ray powder diffraction (SXRPD) patterns for DMAPbI$_3$. (right) Temperature dependence of the unit cell parameters in both LT- and HT-phases. The dotted line indicates the phase transition temperature.

From the obtained structural data, the linear thermal expansion (TE) coefficients ($\alpha$) and the volumetric thermal expansion (TE) coefficient ($\beta_V$) were calculated using the web-based tool PASCal. V2.2.0.[37] For the LT-phase, the following values were obtained: $\alpha_{[-101]}$= -30(7)×10$^{-6}$ K$^{-1}$, $\alpha_b$= 113(6)×10$^{-6}$ K$^{-1}$, $\alpha_{[101]}$= 97(6)×10$^{-6}$ K$^{-1}$ and $\beta_V$= 181(4)×10$^{-6}$ K$^{-1}$. Meanwhile, for the HT-phase, the obtained values were: $\alpha_a$= 95(1) ×10$^{-6}$ K$^{-1}$, $\alpha_c$= -23(1) ×10$^{-6}$ K$^{-1}$ and $\beta_V$= 168×10$^{-6}$ K$^{-1}$. Remarkably, this compound exhibits an anomalous thermomechanical response with negative uniaxial thermal expansion in both phases across the entire studied temperature range. It is worth noting that this compound exhibits a 1D packing, in which the anomalous TE are scarce. [38]

In this context, the direction of negative thermal expansion (TE) differs between the LT- and HT-phases: for the LT-phase, the direction with negative TE is perpendicular to the infinite chains of face-sharing [PbI$_6$]$^{4-}$ octahedra, while in the HT-phase, it is parallel to the chains. Therefore, the origin of the anomalous thermal expansion (TE) differs in both phases. In the case of the LT-phase, it exhibits a monoclinic symmetry in which the beta angle decreases upon heating (see Figure 2), causing expansion along the [101] direction and shrinkage along the perpendicular [-101] direction, see Figure S3 of S.I. Observing the crystal structure, we can deduce that the infinite chains of face-sharing [PbI$_6$]$^{4-}$ octahedra are slightly tilted out of perpendicular alignment to the *a*-axis, and this tilting gradually decreases upon heating until they reach full perpendicularity in the HT-phase. Therefore, this cooperative rotation of all chains is responsible for the observed thermomechanical effect at LT-phase.

Unlike, the HT-phase exhibits a shrinkage along the infinite chains of face-sharing [PbI$_6$]$^{4-}$ octahedra. To gain information about this thermomechanical response, we analyzed the temperature dependence of the Pb-I bond length and Pb-Pb distances, as shown in Figure 3. We observed that in the LT-phase, the Pb$^{2+}$ cations are bonded to I$^-$ anions with six different bond lengths. The average Pb-I bond length decreases slightly, while the Pb-Pb distance increases upon heating over the studied temperature range. In the LT-phase, the infinite chains of face-sharing [PbI$_6$]$^{4-}$ octahedra exhibit positive thermal expansion. In contrast, both Pb-I bond length and Pb-Pb distances decreases upon heating at the HT-phase showing a negative TE. We suggest that this shortening of Pb-I bond lengths and Pb-Pb distances upon heating can be associated with the ionic mobility of I⁻ anions, such behavior has been previously reported in this compound.[39]

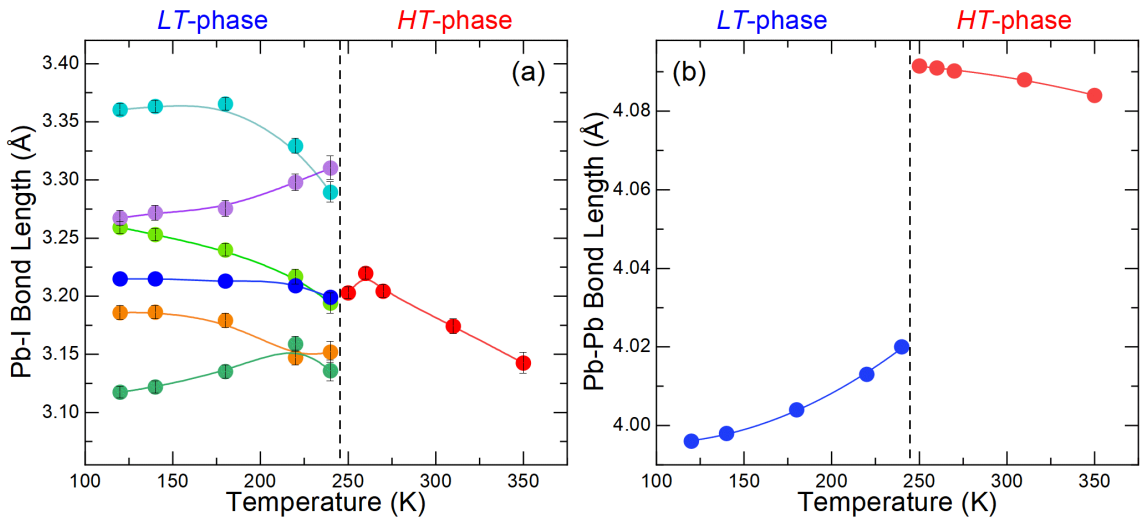

**Figure 3** – (left) Temperature dependence of Pb-I bond length both LT- and HT-phases. Blue square shows the average value of the six different Pb-I bond length at LT-phase. (right) Temperature dependence of Pb-Pb distance for both LT- and HT-phases. The dotted line indicates the phase transition temperature.

Figure 4 presents the temperature-dependent Raman spectra of DMAPbI$_3$. The spectra show a broad range of bands, covering frequencies from 20 to 3100 cm$^{-1}$, which resembles the spectrum of DMAPbBr$_3$.[29] Hence, the spectrum obtained at 300 K (HT-phase) features broader bands, especially at lower wavenumbers, which can be attributed to lattice modes. This broadening is a consequence of the inherent disorder present in the high-temperature structure.[31,44] Conversely, vibrational modes are more well-defined at low temperatures, with new modes emerging. It is worth noting that the vibrational modes of DMAPbI$_3$ are situated at lower frequencies than DMAPbBr$_3$, as typically observed in other perovskites or perovskites-like hybrid compounds.[45–48] Since the DMAPbI$_3$ structure is highly disordered in the HT-phase, we have only presented the group theory analysis for the *LT-phase*.[23] Thus, at low temperatures, according to the irreducible representations of the $C_{2h}$ (2/c) group factor and the occupation of Wyckoff sites in the $C_{2h}^5$ ($P2_1/c$) space group, it is predicted 90 Raman active modes ($\Gamma$ = 45 $A_g$ $\oplus$ 45 $B_g$). Detailed information regarding the irreducible representations for each site in the mode distribution for the DMAPbI$_3$ is available in the Supporting Information (Table S1). As centrosymmetric, the *LT-phase* crystal structure exhibits a mutually exclusive relationship between irreducible representations, implying that they are unable to be simultaneously Raman and infrared active. For the *LT-phase*, we have observed 30 Raman-active modes, while at high-temperatures 21 optical phonons were observed. [33,42]

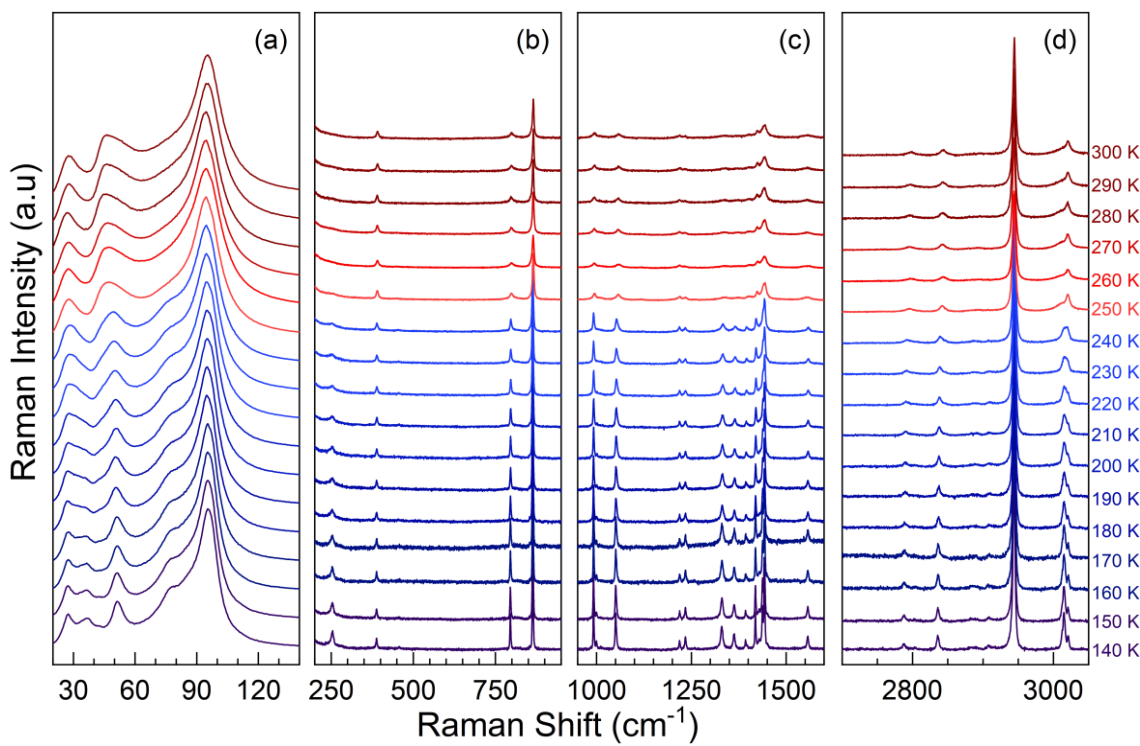

**Figure 4 -** Temperature-dependent normalized Raman spectra obtained for DMAPbI$_3$ in the range between 140 and 300 K in the (a) 20-150 cm$^{-1}$, (b) 200-950 cm$^{-1}$, (c) 950-1600 cm$^{-1}$ and (d) 2700-3050 cm$^{-1}$ range.

To assign these observed modes in both crystalline phases, we employed a comparative approach by aligning our spectra with similar compounds (see Table S2 in the Supporting Information).[45,47,49–53] Notably, we mainly considered the organic-inorganic compound DMAPbBr$_3$.[29] Furthermore, our analysis of temperature-dependent Raman spectra provides valuable insights into the bonding interactions and structural dynamics within the DMAPbI$_3$ crystal lattice. Analysis of wavenumber shifts and bandwidth variations of Raman modes allow us to understand the mechanisms behind phase transitions and order-disorder effects common in these compounds, closely related to their physical properties.

Figure 5 shows the temperature dependence of the Raman-active mode positions of DMAPbI$_3$, divided into distinct regions for easier interpretation. First, our analysis focused on the wavenumber region below 200 cm$^{-1}$ (Figure 5a), which covers the I-Pb-I modes (20-100 cm$^{-1}$) and the rigid-body motion of DMA organic cation modes (100-150

cm$^{-1}$). The I-Pb-I modes are notably influenced by the off-center shift of the lead atoms, which are primarily attributed to the presence of 6s$^2$ electrons *("lone pairs")* within the compound. [54] Despite their chemical inactivity, these lone pairs exert steric effects, leading to an asymmetry in metal coordination and distortion of the [PbI$_6$] octahedron. [55,56] As a result, the bending of I-Pb-I at approximately 26, 45 and 51 cm$^{-1}$ experience a shift in the frequency at the critical temperature, $T_C$ = 250 K. Conversely, the vibrational modes at 35 and 66 cm$^{-1}$ exhibit a sudden shift, indicative of the regularity of the octahedra during the phase transition.[57] In the spectral range associated with the stretching of I-Pb-I, a Raman mode at 95 cm$^{-1}$ splits into two new modes at 96 and 104 cm$^{-1}$ in the low-temperature phase, marking the onset of a structural phase transition at $T_C$. Finally, the mode at around 136 cm$^{-1}$, associated with DMA$^+$ motion, undergoes significant shifts in the high-temperature phase, highlighting the importance of the DMA$^+$ cation on the SPT.

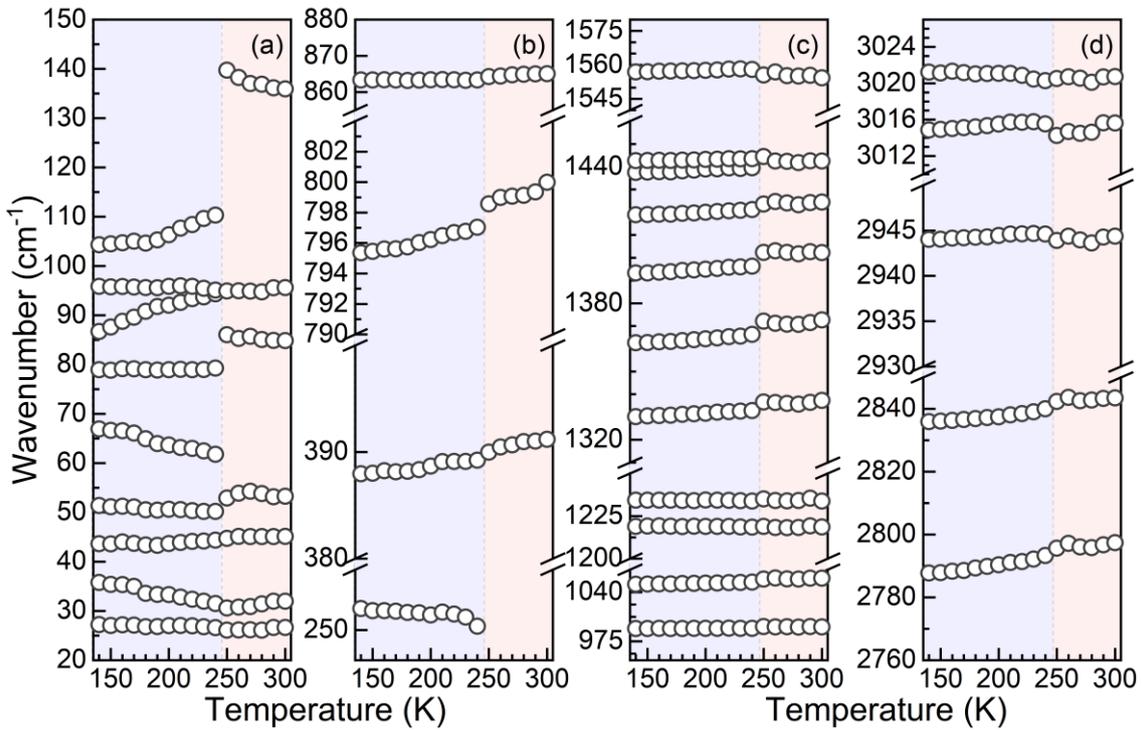

**Figure 5** - Temperature-dependent Raman mode positions obtained for DMAPbI$_3$ in wavenumber ranges: (a) 20-150 cm$^{-1}$, (b) 240-880 cm$^{-1}$, (c) 970-1580 cm$^{-1}$, (d) 2760-3030 cm$^{-1}$.

The Temperature-dependent Raman mode positions in the range 200-950 cm$^{-1}$, which are shown in Figure 5b, can be mainly associated with the torsional motions of the DMA$^+$ molecule (200-450 cm$^{-1}$). In this spectral region, low-intensity modes are observed at 253 and 388 cm$^{-1}$, being the first ones detectable only below 250 K. Furthermore, the modes observed at 795 and 863 cm$^{-1}$ can be assigned as ρ (H$_3$C-N-CH$_3$) rocking and δ (H$_3$C-N-CH$_3$) bending, respectively. Notably, the mode observed at 795 cm$^{-1}$ exhibits a significant redshift at $T_C$ while the mode at 863 cm$^{-1}$ exhibits a subtle redshift at $T_C$, indicating the SPT. These differences may be associated with the presence of two distinct hydrogen bonds in the LT phase: a stronger and linear N – H···I hydrogen bond (d$_{N-I}$ ~3.11 Å) and a weaker, bifurcated one (d$_{N-I}$ ~3.25 Å and d$_{N-I}$ ~3.32 Å).[23]

In the wavenumber range of 950 to 1600 cm$^{-1}$ (Figure 5c), the modes are primarily associated with the complete DMA$^+$ ν (H$_3$C-N-CH$_3$) bending at 992 cm$^{-1}$, along with the bending, stretching, and rocking of the (CH$_3$) and (NH$_2$) groups. Within this range, all the modes exhibit discontinuities at $T_c$. Notably, the modes observed at 1360 cm$^{-1}$ and 1418 cm$^{-1}$ display a significant redshift, along with the splitting of the modes 1437 and 1442 cm$^{-1}$ from the mode at 1400 cm$^{-1}$. As was previously observed in halide hybrid perovskites,[58] it can be inferred that the (H$_3$C-N-CH$_3$) and NH$_2$ bendings are supposed to be sensitive to the hydrogen bond between the DMA cation and the [PbI$_6$] chains, which describes a sudden increase in the full width at half maximum (FWHM), as we will further discuss. These observations highlight the strong correlation between [PbI$_6$] chains and *H-bonds* in stabilizing the DMAPbI$_3$ structure,[59–62] supporting them as indicators of the coupling interaction between them.[63–65]

Finally, the wavenumber range (2700-3050 cm$^{-1}$) displayed in Figure 5d is primarily characterized by CH$_3$ and NH$_2$ stretching modes. The most intense mode observed at 2943 cm$^{-1}$, which can be assigned as ν$_{as}$ (CH$_3$), does not show significant frequency variations within this temperature range, only slightly shifting towards lower wavenumbers. However, a discontinuity in the intensity of this band is observed at $T_c$ due to a sudden increase in its linewidth (not shown here). The main contrast observed in this region is the redshift of the modes observed at 2797 and 2843 cm$^{-1}$ and a subtle blueshift of the modes at 3014 and 3021 cm$^{-1}$, associated with C-H stretching.

On the other hand, considering the dependence of phonon lifetimes on structural characteristics, they are quite relevant in the investigation of the SPT in DMAPbI$_3$ 2H-perovskite. Commonly, important variations of the FWHM can occur with increasing temperature due to anharmonicity and thermally induced reorientation processes.[66] However, within metal-organic frameworks and/or hybrids perovskites, structural order-disorder processes tend to be quite extensive during SPT, and the guest molecules may exhibit highly dynamic behavior. Consequently, at the critical temperature, specific modes are expected to exhibit significant changes in FWHM, associated with the SPT. Additionally, it is important to note that FWHM responds to short- and long-range disorder effects.

As observed in Figure 6 there is a significant discontinuous increase in the FWHM of the selected modes around $T_C$, indicating a shorter phonon lifetime resulting from a highly disordered structure at low temperatures. This discontinuity effect in the linewidth has been extensively observed in hybrid inorganic perovskite, including MA, DMA, and TMA-based compounds. [29,45,46,67] In the short-range, the disorder effects are attributed to the dynamic nature of the octahedra and $H_3C$-N-$CH_3$ skeleton. Thus, our results reveal that the linewidths of the I-Pb-I modes are sensitive to SPT, exhibiting abrupt discontinuity. For instance, the $\nu_s$ (I-Pb-I) vibration at 78 cm$^{-1}$ shows a linewidth shift of approximately 23 cm$^{-1}$, the largest variation among the modes observed in $DMAPbI_3$. This effect can be mainly attributed to the off-center shift of the $Pb^{2+}$ cations, resulting in a large distortion of [$PbI_6$] octahedra, also observed in the $\delta_s$ (I-Pb-I) vibration at 51 cm$^{-1}$. The modes associated with $DMA^+$ skeleton deformation, observed at 795 and 1050 cm$^{-1}$, also exhibit sudden variations in width of approximately 10 and 8 cm$^{-1}$, respectively. This significant broadening of these modes suggests a correlation between the *H*-bond strength and the free motion of the DMA atoms, as well as reaffirms the two cooperative processes around the SPT. [29–31,68]

The broadening of bands in molecular crystals, mainly when influenced by long-range disorder, can be primarily attributed to a combined effect of anharmonicity and the reorientation of molecules or ions.[66] In cases where dynamic disorder prevails over the phonon decay process, the linewidth of a band is typically described using the following equation:

$$FWHM = A + BT + Ce^{(-E_a/kT)} \quad (1)$$

where $E_a$ is the activation energy, $k$ is the Boltzmann constant, and $T$ is the temperature. The constant $A$ considers broadening contributions arising from factors beyond phonon decay, including structural and compositional defects within the crystal lattice. The second ($B$) and third ($C$) terms of Eq. (1) correspond to the effects of anharmonicity and thermally activated reorientation processes, respectively. The anharmonicity term characterizes the deviation from ideal harmonic vibrations in molecular crystals, explaining the intricate behavior of vibrational modes due to intermolecular interactions.

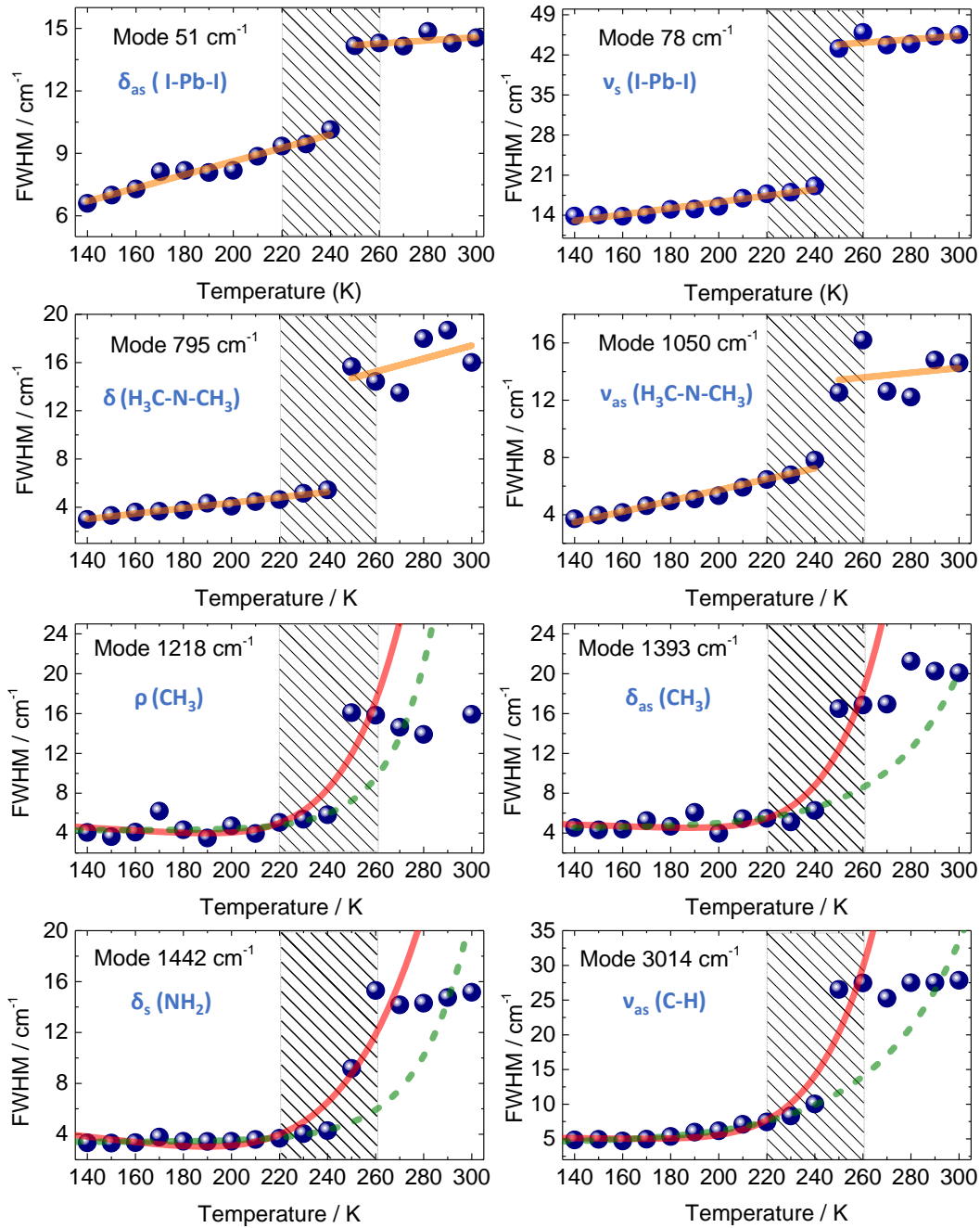

**Figure 6 -** Temperature dependence on full width at half maximum (FWHM) of selected Raman modes of DMAPbI$_3$. The solid orange lines represent a guide for the eyes. Solid red lines are fitted to the data using Eq. (1). The dashed green lines represent the temperature dependence of FWHM in the absence of an SPT. The shaded area represents the region where predominantly DMA reorientation processes occur.

The term associated with thermally activated reorientation processes describes the rotational motion of molecules or ions within the crystal lattice, driven by thermal energy. This model has been effectively utilized to predict the activation energy in organic-inorganic hybrid compounds undergoing order-disorder phase transitions.[69,70] Therefore, Eq. (1) enables us to determine the activation energy related to the reorientation processes of the DMA$^+$ cation in the DMAPbI$_3$ compound. Then, the activation energy was estimated by analyzing the temperature-dependent changes in the FWHM of the Raman-active modes (CH$_3$), $\delta_{as}$ (CH$_3$), $\delta_s$ (NH$_2$), and $\nu_{as}$ (C-H) (see Figure 6). All pertinent information derived from the fitting was concisely summarized in Table S3 (see S.I).

As shown in Table S3 some selected modes from the DMA, such as those observed at 1218 and 1393 cm$^{-1}$, display a significant broadening of about 10 cm$^{-1}$. The fit by Eq. (1) yield activation energies of 0.27 eV (KJ/mol) for $\rho$ (CH$_3$) and $E_a = 0.30$ meV (KJ/mol) for $\delta_{as}$ (CH$_3$). Conversely, the modes identified at 1442 and 3014 cm$^{-1}$ exhibit $E_a = 0.21$ eV for $\delta_s$ (NH$_2$) and $E_a = 0.26$ eV for $\nu_{as}$ (C-H), respectively. These findings indicate that the linear terms ($B$ values) are notably lower for the modes detected at 1393 and 3014 cm$^{-1}$, suggesting a diminished anharmonic contribution. Moreover, our study allows a comparison of the potential barrier values obtained for other DMA-based compounds. For instance, the compound [(CH$_3$)$_2$NH$_2$]$_3$[Bi$_2$Cl$_9$] displayed the $E_a$ value of 0.19 eV, [71] while [(CH$_3$)$_2$NH$_2$]$_2$KCr(CN)$_6$ exhibited an $E_a$ value of 0.23 eV.[72] Similarly, the compounds as [(CH$_3$)$_2$NH$_2$]$_5$Cd$_3$Cl$_{11}$ ($E_a = 0.19$ eV), [(CH$_3$)$_2$NH$_2$][Zn(HCOO)$_3$] ($E_a = 0.24$ eV),[73,74] [(CH$_3$)$_2$NH$_2$]Na$_{0.5}$Fe$_{0.5}$(HCOO)$_3$] ($E_a = 0.28$ eV),[75] and [(CH$_3$)$_2$NH$_2$][Cd(N$_3$)$_3$][76] ($E_a = 0.38$ eV). Therefore, it is evident that the activation energy for the DMA cation in DMAPbI$_3$ resembles that of other DMA-based materials.

Since the ability of the guest cation to rotate or hop is inherently linked to the energy barrier, a rotating or hopping state can persist even at very low temperatures if $E_a$ value is low. Indeed, we observed an order-disorder transition at 250 K for DMAPbI$_3$, at lower temperatures than other similar compounds, such as TrMAMnN$_3$ (332 K), which has higher activation energies (0.64 eV)[77]. The activation energy is primarily attributed to the increased mass and larger dimensions of the (CH$_3$)$_3$NH$^+$ cation as a guest species. It's also noteworthy that our analysis unveils that the influence of reorientation processes on SPT becomes significant above approximately 220 K, as evidenced by the deviation of the models for mode widths within the transition region (Figure 6). However, just below 220 K, the contribution related to DMA reorientation processes becomes negligible, and the observed changes in the FWHM are predominantly associated with anharmonicity. Thus, the Raman spectroscopy technique is sensitive for identifying the temperature ranges where the reorientation processes of DMA occur in these structures.

**B. Polar phonons, intrinsic dielectric, and optical room-temperature properties**

Figure 7 shows the FTIR reflectivity spectrum of the DMAPbI$_3$ compound at room temperature. This spectrum has been fitted using the four-parameter semi-quantum model developed by Gervais and Piriou,[40] as detailed in the Supporting Information (S.I). By this model, we can obtain the vibrational characteristics of polar phonons, such as frequencies and damping coefficients. Additionally, this approach provides insights into the intrinsic dielectric responses of the compound, indicating which specific polar phonons contribute more to the intrinsic dielectric constant of the material.

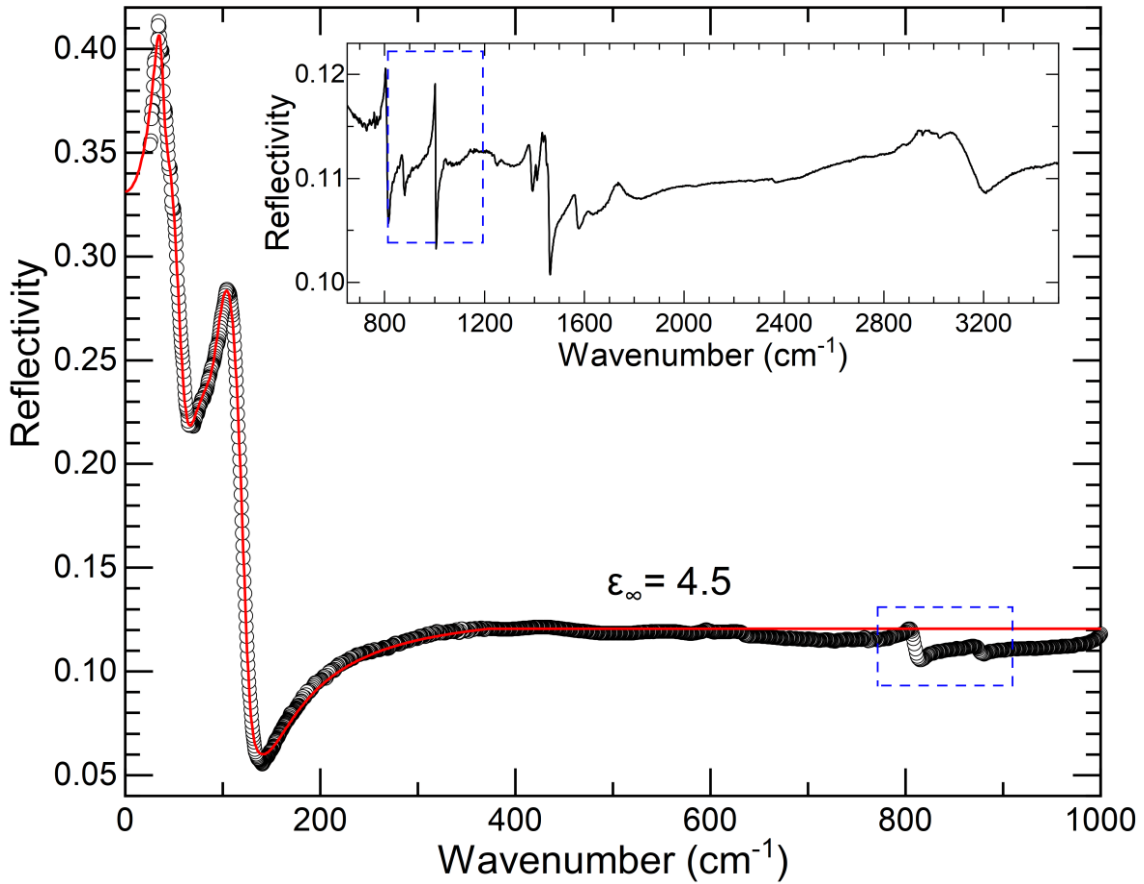

**Figure 7.** Infrared reflectivity spectrum of DMAPbI₃ at room temperature. The red circles represent the experimental data, while the black line curve shows the simulated spectrum using the isotropic dielectric function obtained from the four-parameter semi-quantum model.

The intrinsic dielectric constant ($\varepsilon_s$) was evaluated at approximately 13.7, where the main contributions were from the far-infrared region, up to 600 cm$^{-1}$. As depicted in Figure 7 (inset), the internal modes of the DMA cation, observed beyond 800 cm$^{-1}$, are considerably weak and narrow due to their very low dielectric strengths ($\Delta\varepsilon_j$), *i.e.*, there is a low charge dynamic associated with those modes. In general, these higher frequency modes do not contribute significantly to the intrinsic dielectric constant of the compound, which above 400 cm$^{-1}$ is dominated by electronic polarization, accounting for $\varepsilon_\infty$. For DMAPbI₃, the $\varepsilon_\infty$ value was determined to be 4.5, resulting in a refractive index of $n = 2.12$. This value aligns with previously reported data for similar lead halide perovskites, such as MAPbCl₃ ($\varepsilon_\infty = 4.0$), MAPbBr₃ ($\varepsilon_\infty = 4.7$) and MAPbI₃ ($\varepsilon_\infty = 5.0$).[41]

It is noteworthy that the modes that most contribute to the intrinsic dielectric constant are observed below 150 cm$^{-1}$. Therefore, assigning the vibrational characteristics of these modes becomes crucial for understanding their contribution. The modes observed within the 20-100 cm$^{-1}$ range are associated with the stretching and bending vibration of the I-Pb-I bonds. Particularly, considering the longitudinal frequencies ($\omega_{j,LO}$), we observed that the lowest frequency modes observed at 39 and 59 cm$^{-1}$ exhibit a robust dielectric strength, evidenced by its $\Delta\varepsilon_j$ value of 2.77 and 4.05, respectively. Excluding the electronic polarization, these modes collectively account for approximately 74% of the static (infrared) dielectric constant and are responsible for generating the most significant dipole moment within the spectrum of DMAPbI₃. Table S4 shows the complete phonon and dielectric parameters obtained from the fitting.

Conversely, the modes observed in the 100 to 128 cm$^{-1}$ range, attributed to DMA motion (translation or libration), demonstrate comparatively lower dielectric strength contribution. Notably, this observation aligns with findings for MAPbI₃, where FTIR reflectivity measurements indicated minimal contributions from MA$^+$ modes to the intrinsic dielectric constant.[42] Furthermore, it is worth noting that the intrinsic dielectric constant value of 13.72 at room temperature indicates that the previously reported dielectric constant for bulk DMAPbI$_3$ ($\varepsilon'_r$~1000, at 325 K for $\nu$ = 10 Hz) mainly arises from extrinsic processes related to the migration of iodine anions due to interstitial defects or vacancies.[23]

To investigate the optical properties of the DMAPbI$_3$ compound, UV-vis absorption spectra were recorded on a solid powder sample pellet. The material displayed an absorbance within the range of 450 to 700 nm (see **Figure S5(a)**). The sample bandgap ($E_g$) value was calculated via a Tauc plot method. [43] The determined $E_g$, as shown in **Figure S5 (b)**, indicates a direct semiconductor-type behavior. The calculated room-temperature optical bandgap for DMAPbI$_3$ was $E_g$ = 2.31 eV, which is consistent with the previous values reported in the literature. [31,44]

## C. Structural phase transition and optoelectronic properties in DMAPbI$_3$

Since we investigated in detail the SPT observed in DMAPbI$_3$ at low temperatures as well as determined the optical properties at room temperature, we can investigate the SPT relationship with the optoelectronic properties at low temperatures. Thus, we investigate the photoluminescent properties at low temperatures to gain insights into the origin of the broadband light emission of DMAPbI$_3$. Figure 8 presents a high-resolution intensity map scale generated from DMAPbI$_3$ when excited using white LED light (Figure 8a) and UV 405 nm light (Figure 8b) in the temperature range from 300 down to 170 K. Under white light irradiation, the crystal exhibits a thermochromic effect, where the color of the material shifts from a yellowish-white to a white color (Figure 8a). Then, when exposed to UV light it emits a vivid orange-red photoluminescence (PL) at low temperatures (Figure 8b). Both optical phenomena become more evident after the complete transformation from *HT-phase* to *LT-phase*, just after the $T_c$ as presented in the cooling DSC measurement (Figure 8c).

As observed in Figure 8a, the thermochromic effect intensity increases with decreasing temperature in DMAPbI$_3$, similar to the behavior observed in MAPbI$_3$ [78], in which the thermochromism arises from a reversible hydration/dehydration process induced by its SPT. [79,80] However, in DMAPbI$_3$, the thermochromic effect was observed both when the system was cooled in air and in an evacuated system, indicating that the observed thermochromism is not associated with the same mechanism as MAPbI$_3$.

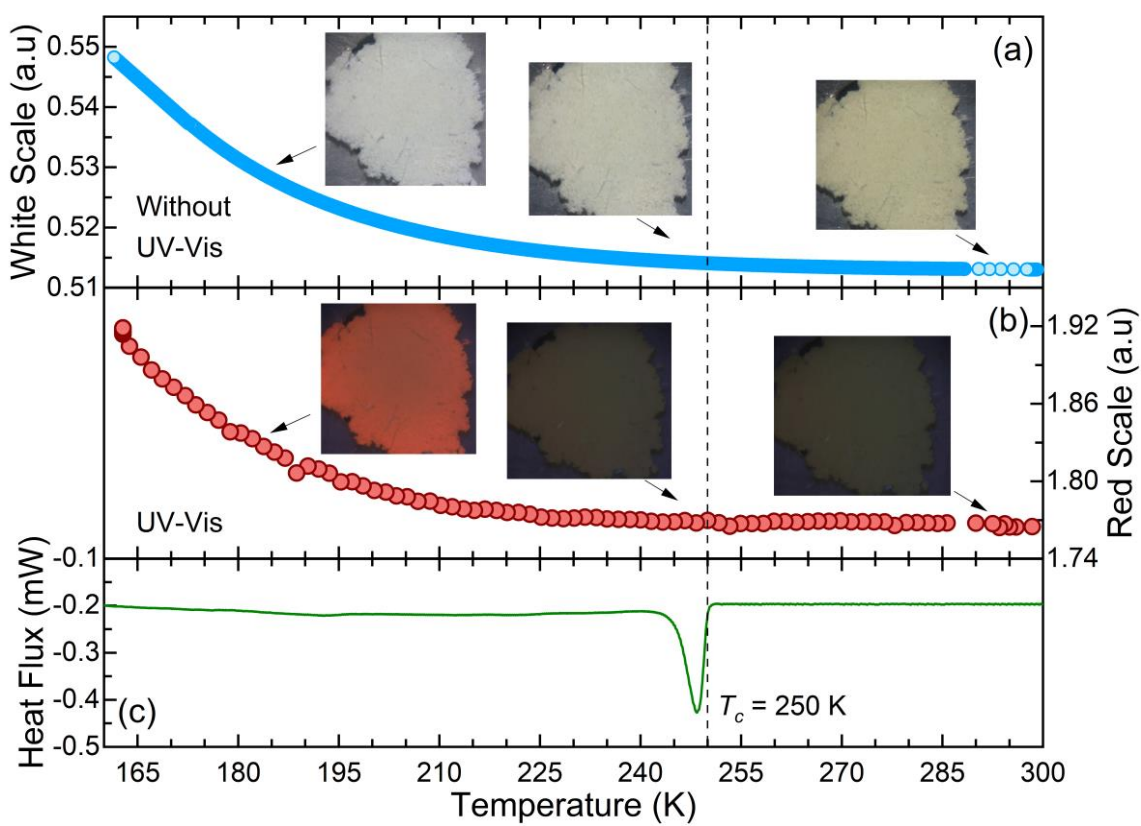

**Figure 8** – High-resolution intensity maps of DMAPbI$_3$ showing **(a)** the thermochromic effects and **(b)** the light orange-red photoluminescence after the structural phase transition, accompanied by **(c)** the cooling DSC process. The orange dashed bar separates the two phases, indicating the trend modification of both curves around T$_c$.

Figure 8a displays the temperature-dependent photoluminescence (PL) observed in the temperature range from 300 down to 10K for DMAPbI$_3$. At room temperature no photoluminescence (PL) emission is observed. However, just below $T_C$, the compound exhibits a light broadband emission centered around 680 nm (1.82 eV), in which the intensity increases with decreasing temperature. This broadband emission resembles that observed in other related one-dimensional perovskites, which is attributed to a self-trapped exciton (STE) mechanism. [81–83] Figure 8b shows the temperature dependence of the center of the broadband emission. It is well-established in the current literature that the formation of STE presumes a strong electron-phonon coupling, which generally can be evaluated by the well-known Huang−Rhys factor $S$.[84] The electron-phonon coupling reveals a strong connection with the FWHM of the luminescence band, as described by the following equation: [85,86]

$$FWHM(T) = 2.36\sqrt{S}\hbar\omega \sqrt{\coth\frac{\hbar\omega}{2k_bT}} \qquad (2)$$

where the $S$ is the Huang-Rhys factor, and $\hbar\omega$ is the average optical phonon energy. Figure 8c describes the FWHM temperature dependence of PL.

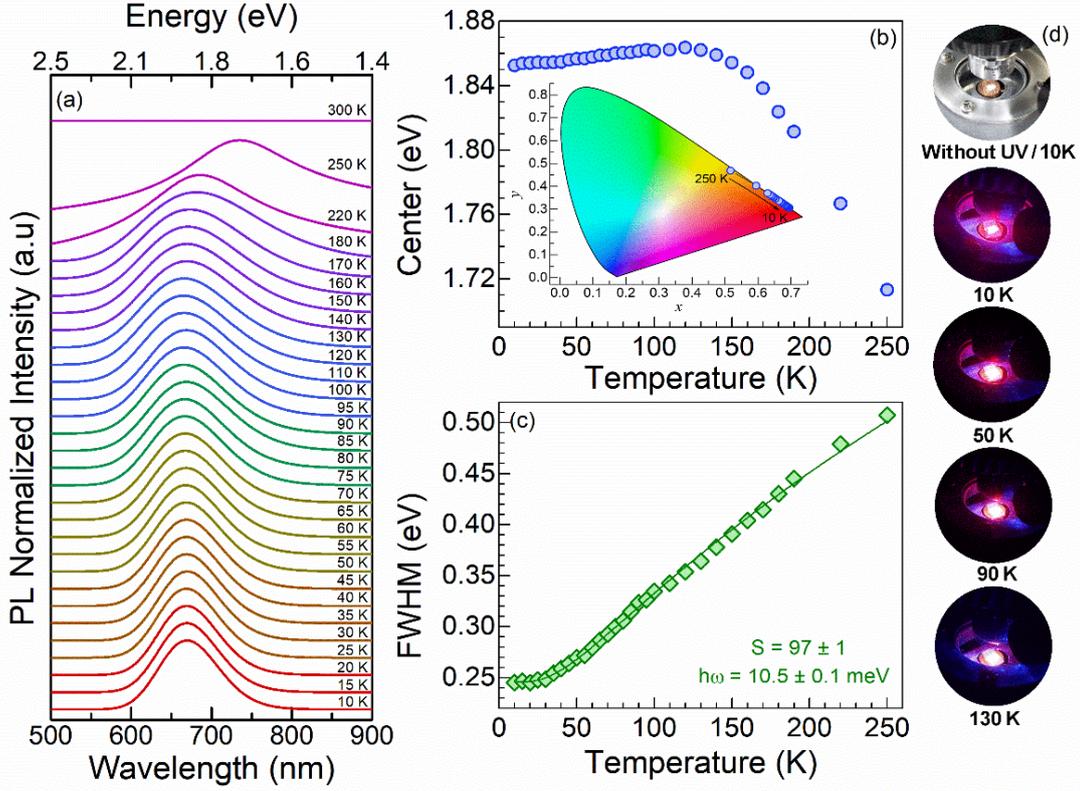

**Figure 8** – (a) Temperature-dependence PL spectra of DMAPbI$_3$, obtained by exciting the sample with 405 nm light. (b) The low-temperature dependence of peak centers of PL band. The inset shows the Temperature-dependent chromaticity coordinates (CIE) of the emissions. (c) The FWHM temperature-dependence of the PL band where the data fits Eq (2), and (d) Photograph of DMAPbI$_3$ sample under ultraviolet (UV) light at different temperatures and without UV applied.

The increasing intensity and presence of STE energy level in DMAPbI$_3$ may provide insight into the electron-phonon coupling strength of the material. Thus, the model described in Eq. (2) is utilized primarily to estimate the Huang-Rhys factor according to the temperature dependence of FWHM of the STE photoluminescence band. Therefore, the S value defines the electron-phonon coupling strength, playing a crucial role in evaluating and optimizing the emission characteristics of STEs and elucidating carrier trapping mechanisms in soft semiconductors (Usually, $S \gg 1$ indicates a strong coupling regime).[87–89] In DMAPbI$_3$, the estimated values of $S = 97 \pm 1$, imply a strong electron-phonon coupling regime, much higher compared to other halide perovskites such as the inorganics CsPbBr$_3$ ($S = 12$) [90], Cs$_3$Sb$_2$I$_9$ ($S = 42.7$), Rb$_3$Sb$_2$I$_9$ ($S = 50.4$), Cs$_3$Bi$_2$I$_9$, ($S = 79.5$) [91] and 1D hybrids perovskites as C$_5$H$_{16}$N$_2$Pb$_2$Br$_6$ ($S = 54$), [C$_6$H$_7$ClN]CdCl$_3$ ($S = 62.94$), [92] and [C$_7$H$_{10}$N]$_3$[BiCl$_5$]Cl ($S = 45$). [93] On the other hand, average optical phonon energy $\hbar\omega$ establishes the effective frequency value of the LO mode that contributes most significantly to the electron-phonon coupling. Thus, the calculated value for DMAPbI$_3$ was $\hbar\omega = 10.4 \pm 0.1\ meV \sim 84\ cm^{-1}$, which agrees well with the frequency region observed by FTIR, as previously presented in Table S4.

Therefore, the low-temperature photoluminescence phenomenon observed in DMAPbI$_3$ can be attributed to the self-trapped excitons (STEs) produced by photogenerated electron-holes which are allowed by crystalline distortions driven by the SPT.[94,95] The difference in photoluminescence between the HT and LT phases can be explained by considering thermal lattice and electron-phonon coupling. [96] Thermal effects at higher temperatures contribute to non-radiative pathways for excited charge carriers. This minimizes the contribution of electron-phonon coupling to luminescence emission. This is called thermal quenching and is well-established in the literature.[97–99] However, at low temperatures, due to the structural phase transition, the dynamics of inorganic off-center displacement of the Pb$^{+2}$ ion produce distortions in the octahedra. These distortions are followed by electron capture, favoring the strong electron-phonon coupling and producing self-trapped excitons.

In addition to the strong electron-phonon coupling (S = 97 ± 1), according to Ying Han *et al.*, there are other factors influencing self-trapped exciton emission of low-dimensional metal halides.[100] Firstly, the width of PL emission can be associated with the distortion degree of [PbX$_6$] octahedra.[100] It has been observed that larger octahedral distortion levels in 2D lead bromide perovskites exhibit a linear dependence on the broader width of PL emission.[101–103] A similar approach can be considered to evaluate the octahedral distortion in 1D halide systems.[104] Thus, considering the structural data of DMAPbI$_3$ at 100 K (*LT-phase*) and 295 K (*HT-phase*) obtained from the reference,[23] it is possible to evaluate the distortion of [PbI$_6$] octahedron arising from the variation of the Pb−I bond length by the following equation[85,86]

$$\Delta d = \left(\frac{1}{6}\right) \sum \left[\frac{d_n - d}{d}\right]^2 \quad (2)$$

where $d_n$ is the individual Pb−I bond length and $d$ represents the mean Pb – I distance. In *HT-phase* the parameter $\Delta d$ is zero as the octahedra [PbI$_6$] are regular. Otherwise, for the *LT- phase* the obtained value reveals that octahedral chains have are highly distorted ($\Delta d$ = 5.59 10$^{-4}$). This octahedral distortion degree observed is comparable to those of 1D organic lead halide perovskites, [104–106] providing insights into the nature of its STE broadband light emission. [100,104,107]

Secondly, another factor that plays a crucial role in STE formation is attributed to the shortening of the Pb–Pb bonds due to lattice distortion.[100] We observe that the Pb–Pb bonds in the *LT- phase* (d$_{Pb-Pb}$ ~3.992 Å) are shortened compared to the *HT- phase* (d$_{Pb-Pb}$ ~4.057 Å). When the Pb–Pb bonds are shortened, photogenerated electrons are more easily captured at the Pb–Pb sites, and these electrons couple with the Pb–Pb dimers, generating Pb$_2^{3+}$ centers, i.e., the formation of small polarons located at [PbI$_6$]$^{2-}$ octahedra which strongly agree with the proposed STE with high Huang-Rhys factor S.[100,108] Consequently, intermediate energy self-trapped exciton states are introduced into the forbidden band, facilitating the trapping of electrons and holes.[100,109]

To investigate the potential applications of intense red light-emitting diodes (LEDs), the effect of spectral power distribution was evaluated by the CIE 1931 color space curve, which describes the (*x,y*) chromaticity coordinates.[110] The chromaticity coordinates of DMAPbI$_3$ in the CIE-1931 color space as a function of temperature are presented in the inset of Figure 8b, with the corresponding values given in Table S5. There is a distinct color transition from red-orange at 180 K to intense bright red at lower temperatures, as depicted in Figure 8c.

## IV. CONCLUSIONS

We investigated the thermal evolution of crystal structure, lattice dynamics and optoelectronic properties of [(CH$_3$)$_2$NH$_2$]PbI$_3$ hybrid perovskite. Using VT-SXRPD experiments, we have obtained valuable information about the thermal evolution of both LT- and HT-crystal structure close to the phase transition temperature. Notably, the Pb-I bond lengths shorten upon heating at both phases and the Pb-Pb distance increases with temperature in the LT-phase and decreases in the HT-phase. Additionally, this compound exhibits uniaxial negative thermal expansion (TE) in the temperature range of 120-350 K. Room temperature FTIR reflectivity spectra identified polar phonons and their damping coefficients. Minimal contributions from the $DMA^+$ mode were observed in the intrinsic dielectric constant (~13.72), with the stretching and bending modes of I-Pb-I exhibiting the most significant contributions. The optical properties of $DMAPbI_3$ were characterized by an optical bandgap of $E_g$ = 2.31 eV, indicating a direct semiconductor-type behavior. Additionally, Raman Spectroscopy was employed to study the structural phase transition in $DMAPbI_3$, revealing that the modes of I-Pb-I and $DMA^+$ display an important role in the order-disorder transition in this material. Phonon lifetime analysis investigated short and long-range disorder effects, revealing the impact of $DMA^+$ reorientation processes on the observed discontinuous width broadening. This study underscores the effectiveness of Raman spectroscopy in identifying temperature ranges conducive to $DMA^+$ reorientational processes within these structures. Additionally, low-temperature measurements were conducted to probe the optoelectronic properties of $DMAPbI_3$. The low-temperature photoluminescence phenomenon has been associated with the presence of self-trapped excitons (STEs), which arise from photogenerated holes due to the high octahedral distortion during the SPT. The presence of the STE band in $DMAPbI_3$ is evidenced by the strong electron-phonon coupling, confirmed by a calculated Huang-Rhys factor of S = 97 ± 1. Thus, we have addressed the light broadband emission in $DMAPbI_3$ as a combination of factors: strong spin-phonon coupling, high octahedral distortion, and the shortening of Pb-Pb bonds. Moreover, the effective calculated frequency value of the LO mode, which significantly contributes to electron-phonon coupling (84 cm$^{-1}$), aligns well with the findings obtained from FTIR analysis. Finally, these results underscore the intricate interplay between electronic, SPT and vibrational properties in $DMAPbI_3$, shedding light on its fundamental physics and paving the way for further exploration and understanding of its optoelectronic features.

## V. REFERENCES

(1) Liu, X.-K.; Xu, W.; Bai, S.; Jin, Y.; Wang, J.; Friend, R. H.; Gao, F. Metal Halide Perovskites for Light-Emitting Diodes. *Nat Mater* **2021**, *20* (1), 10–21. https://doi.org/10.1038/s41563-020-0784-7.

(2) Bresolin, B.-M.; Park, Y.; Bahnemann, D. Recent Progresses on Metal Halide Perovskite-Based Material as Potential Photocatalyst. *Catalysts* **2020**, *10* (6), 709. https://doi.org/10.3390/catal10060709.

(3) Schmidt-Mende, L.; Dyakonov, V.; Olthof, S.; Ünlü, F.; Lê, K. M. T.; Mathur, S.; Karabanov, A. D.; Lupascu, D. C.; Herz, L. M.; Hinderhofer, A.; Schreiber, F.; Chernikov, A.; Egger, D. A.; Shargaieva, O.; Cocchi, C.; Unger, E.; Saliba, M.; Byranvand, M. M.; Kroll, M.; Nehm, F.; Leo, K.; Redinger, A.; Höcker, J.; Kirchartz, T.; Warby, J.; Gutierrez-Partida, E.; Neher, D.; Stolterfoht, M.; Würfel, U.; Unmüssig, M.; Herterich, J.; Baretzky, C.; Mohanraj, J.; Thelakkat, M.; Maheu, C.; Jaegermann, W.; Mayer, T.; Rieger, J.; Fauster, T.; Niesner, D.; Yang, F.; Albrecht, S.; Riedl, T.; Fakharuddin, A.; Vasilopoulou, M.; Vaynzof, Y.; Moia, D.; Maier, J.; Franckevičius, M.; Gulbinas, V.; Kerner, R. A.; Zhao, L.; Rand, B. P.; Glück, N.; Bein, T.; Matteocci, F.; Castriotta, L. A.; Di Carlo, A.; Scheffler, M.; Draxl, C. Roadmap on Organic-Inorganic Hybrid Perovskite Semiconductors and Devices. *APL Mater* **2021**, *9* (10). https://doi.org/10.1063/5.0047616/123056.

(4) Guan, X.; Lei, Z.; Yu, X.; Lin, C.; Huang, J.; Huang, C.; Hu, L.; Li, F.; Vinu, A.; Yi, J.; Wu, T. Low-Dimensional Metal-Halide Perovskites as High-Performance Materials for Memory Applications. *Small* **2022**, *18* (38). https://doi.org/10.1002/smll.202203311.

(5) Lin, K.; Xing, J.; Quan, L. N.; de Arquer, F. P. G.; Gong, X.; Lu, J.; Xie, L.; Zhao, W.; Zhang, D.; Yan, C.; Li, W.; Liu, X.; Lu, Y.; Kirman, J.; Sargent, E. H.; Xiong, Q.; Wei, Z. Perovskite Light-Emitting Diodes with External Quantum Efficiency Exceeding 20 per Cent. *Nature* **2018**, *562* (7726), 245–248. https://doi.org/10.1038/s41586-018-0575-3.

(6) Lin, C.-F.; Huang, K.-W.; Chen, Y.-T.; Hsueh, S.-L.; Li, M.-H.; Chen, P. Perovskite-Based X-Ray Detectors. *Nanomaterials* **2023**, *13* (13), 2024. https://doi.org/10.3390/nano13132024.

(7) Yakunin, S.; Sytnyk, M.; Kriegner, D.; Shrestha, S.; Richter, M.; Matt, G. J.; Azimi, H.; Brabec, C. J.; Stangl, J.; Kovalenko, M. V.; Heiss, W. Detection of X-Ray Photons by Solution-Processed Lead Halide Perovskites. *Nat Photonics* **2015**, *9* (7), 444–449. https://doi.org/10.1038/nphoton.2015.82.

(8) Zhou, J.; Huang, J. Photodetectors Based on Organic-Inorganic Hybrid Lead Halide Perovskites. *Advanced Science* **2018**, *5* (1), 1700256. https://doi.org/10.1002/advs.201700256.

(9) Huang, J.; Lai, M.; Lin, J.; Yang, P. Rich Chemistry in Inorganic Halide Perovskite Nanostructures. *Advanced Materials* **2018**, *30* (48), 1802856. https://doi.org/10.1002/ADMA.201802856.

(10) Mao, L.; Guo, P.; Kepenekian, M.; Hadar, I.; Katan, C.; Even, J.; Schaller, R. D.; Stoumpos, C. C.; Kanatzidis, M. G. Structural Diversity in White-Light-Emitting Hybrid Lead Bromide Perovskites. *J Am Chem Soc* **2018**, *140* (40), 13078–13088. https://doi.org/10.1021/jacs.8b08691.

(11) Lekina, Y.; Shen, Z. X. Excitonic States and Structural Stability in Two-Dimensional Hybrid Organic-Inorganic Perovskites. *Journal of Science: Advanced Materials and Devices* **2019**, *4* (2), 189–200. https://doi.org/10.1016/j.jsamd.2019.03.005.

(12) Akkerman, Q. A.; Manna, L. What Defines a Halide Perovskite? *ACS Energy Lett* **2020**, *5* (2), 604–610. https://doi.org/10.1021/ACSENERGYLETT.0C00039/ASSET/IMAGES/LARGE/NZ0C00039_0005.JPEG.

(13) Gao, Y.; Shi, E.; Deng, S.; Shiring, S. B.; Snaider, J. M.; Liang, C.; Yuan, B.; Song, R.; Janke, S. M.; Liebman-Peláez, A.; Yoo, P.; Zeller, M.; Boudouris, B. W.; Liao, P.; Zhu, C.; Blum, V.; Yu, Y.; Savoie, B. M.; Huang, L.; Dou, L. Molecular Engineering of Organic–Inorganic Hybrid Perovskites Quantum Wells. *Nat Chem* **2019**, *11* (12), 1151–1157. https://doi.org/10.1038/s41557-019-0354-2.

(14) García-Fernández, A.; Juarez-Perez, E. J.; Bermúdez-García, J. M.; Llamas-Saiz, A. L.; Artiaga, R.; López-Beceiro, J. J.; Señarís-Rodríguez, M. A.; Sánchez-Andújar, M.; Castro-García, S. Hybrid Lead Halide [(CH3)2NH2]PbX3 (X = Cl- and Br-) Hexagonal Perovskites with Multiple Functional Properties. *J Mater Chem C Mater* **2019**, *7* (32), 10008–10018. https://doi.org/10.1039/c9tc03543e.

(15) Chen, Q.; De Marco, N.; Yang, Y.; Song, T. Bin; Chen, C. C.; Zhao, H.; Hong, Z.; Zhou, H.; Yang, Y. Under the Spotlight: The Organic–Inorganic Hybrid Halide Perovskite for Optoelectronic Applications. *Nano Today* **2015**, *10* (3), 355–396. https://doi.org/10.1016/J.NANTOD.2015.04.009.

(16) Pedesseau, L.; Sapori, D.; Traore, B.; Robles, R.; Fang, H. H.; Loi, M. A.; Tsai, H.; Nie, W.; Blancon, J. C.; Neukirch, A.; Tretiak, S.; Mohite, A. D.; Katan, C.; Even, J.; Kepenekian, M. Advances and Promises of Layered Halide Hybrid Perovskite Semiconductors. *ACS Nano* **2016**, *10* (11), 9776–9786. https://doi.org/10.1021/ACSNANO.6B05944/ASSET/IMAGES/MEDIUM/NN-2016-059449_0007.GIF.

(17) Yan, J.; Qiu, W.; Wu, G.; Heremans, P.; Chen, H. Recent Progress in 2D/Quasi-2D Layered Metal Halide Perovskites for Solar Cells. *J Mater Chem A Mater* **2018**, *6* (24), 11063–11077. https://doi.org/10.1039/C8TA02288G.

(18) Ishihara, T. OPTICAL PROPERTIES OF Pb -BASED INORGANIC-ORGANIC PEROVSKITES . *Optical Properties of Low–Dimensional Materials* **1996**, 288–339. https://doi.org/10.1142/9789814261388_0006.

(19) Zhou, G.; Su, B.; Huang, J.; Zhang, Q.; Xia, Z. Broad-Band Emission in Metal Halide Perovskites: Mechanism, Materials, and Applications. *Materials Science and Engineering R: Reports*. Elsevier Ltd July 1, 2020. https://doi.org/10.1016/j.mser.2020.100548.

(20) Lin, H.; Zhou, C.; Tian, Y.; Siegrist, T.; Ma, B. Low-Dimensional Organometal Halide Perovskites. *ACS Energy Lett* **2018**, *3* (1), 54–62. https://doi.org/10.1021/acsenergylett.7b00926.

(21) Guo, T. M.; Gao, F. F.; Li, Z. G.; Liu, Y.; Yu, M. H.; Li, W. Mechanical and Acoustic Properties of a Hybrid Organic-Inorganic Perovskite, TMCM-CdCl3, with Large Piezoelectricity. *APL Mater* **2020**, *8* (10), 101106. https://doi.org/10.1063/5.0027776/1062924.

(22) Hu, T.; Smith, M. D.; Dohner, E. R.; Sher, M. J.; Wu, X.; Trinh, M. T.; Fisher, A.; Corbett, J.; Zhu, X. Y.; Karunadasa, H. I.; Lindenberg, A. M. Mechanism for Broadband White-Light Emission from Two-Dimensional (110) Hybrid Perovskites. *Journal of Physical Chemistry Letters* **2016**, *7* (12), 2258–2263. https://doi.org/10.1021/acs.jpclett.6b00793.

(23) García-Fernández, A.; Bermúdez-García, J. M.; Castro-García, S.; Llamas-Saiz, A. L.; Artiaga, R.; López-Beceiro, J.; Hu, S.; Ren, W.; Stroppa, A.; Sánchez-Andújar, M.; Señarís-Rodríguez, M. A. Phase Transition, Dielectric Properties, and Ionic Transport in the [(CH3)2NH2]PbI3 Organic-Inorganic Hybrid with 2H-Hexagonal Perovskite Structure. *Inorg Chem* **2017**, *56* (9), 4918–4927. https://doi.org/10.1021/acs.inorgchem.6b03095.

(24) Fabini, D. H.; Seshadri, R.; Kanatzidis, M. G. The Underappreciated Lone Pair in Halide Perovskites Underpins Their Unusual Properties. *MRS Bull* **2020**, *45* (6), 467–477. https://doi.org/10.1557/MRS.2020.142.

(25) Knutson, J. L.; Martin, J. D.; Mitzi, D. B. Tuning the Band Gap in Hybrid Tin Iodide Perovskite Semiconductors Using Structural Templating. *Inorg Chem* **2005**, *44* (13), 4699–4705. https://doi.org/10.1021/ic050244q.

(26) Bati, A. S. R.; Zhong, Y. L.; Burn, P. L.; Nazeeruddin, M. K.; Shaw, P. E.; Batmunkh, M. Next-Generation Applications for Integrated Perovskite Solar Cells. *Commun Mater* **2023**, *4* (1), 2. https://doi.org/10.1038/s43246-022-00325-4.

(27) Huang, J.; Yuan, Y.; Shao, Y.; Yan, Y. Understanding the Physical Properties of Hybrid Perovskites for Photovoltaic Applications. *Nat Rev Mater* **2017**, *2*. https://doi.org/10.1038/natrevmats.2017.42.

(28) Liu, X.; Wang, Y.; Wang, Y.; Zhao, Y.; Yu, J.; Shan, X.; Tong, Y.; Lian, X.; Wan, X.; Wang, L.; Tian, P.; Kuo, H.-C. Recent Advances in Perovskites-Based Optoelectronics. *Nanotechnol Rev* **2022**, *11* (1), 3063–3094. https://doi.org/10.1515/ntrev-2022-0494.

(29) Rodríguez-Hernández, J. S.; P. Gómez, M. A.; Abreu, D. S.; Nonato, A.; da Silva, R. X.; García-Fernández, A.; Señarís-Rodríguez, M. A.; Sánchez-Andújar, M.; Ayala, A. P.; Paschoal, C. W. A. Uniaxial Negative Thermal Expansion in the [(CH$_3$)$_2$NH$_2$]PbBr$_3$ Hybrid Perovskite. *J Mater Chem C Mater* **2022**, *10* (46), 17567–17576. https://doi.org/10.1039/D2TC02708A.

(30) García-Fernández, A.; Juarez-Perez, E. J.; Bermúdez-García, J. M.; Llamas-Saiz, A. L.; Artiaga, R.; López-Beceiro, J. J.; Señarís-Rodríguez, M. A.; Sánchez-Andújar, M.; Castro-García, S. Hybrid Lead Halide [(CH$_3$)$_2$NH$_2$]PbX$_3$ (X = Cl$^-$ and Br$^-$) Hexagonal Perovskites with Multiple Functional Properties. *J Mater Chem C Mater* **2019**, *7* (32), 10008–10018. https://doi.org/10.1039/C9TC03543E.

(31) García-Fernández, A.; Bermúdez-García, J. M.; Castro-García, S.; Llamas-Saiz, A. L.; Artiaga, R.; López-Beceiro, J.; Hu, S.; Ren, W.; Stroppa, A.; Sánchez-Andújar, M.; Señarís-Rodríguez, M. A. Phase Transition, Dielectric Properties, and Ionic Transport in the [(CH3)2NH2]PbI3 Organic-Inorganic Hybrid with 2H-Hexagonal Perovskite Structure. *Inorg Chem* **2017**, *56* (9), 4918–4927. https://doi.org/10.1021/acs.inorgchem.6b03095.

(32) Altomare, A.; Cuocci, C.; Giacovazzo, C.; Moliterni, A.; Rizzi, R.; Corriero, N.; Falcicchio, A. *EXPO2013*: A Kit of Tools for Phasing Crystal Structures from Powder Data. *J Appl Crystallogr* **2013**, *46* (4), 1231–1235. https://doi.org/10.1107/S0021889813013113.

(33) Toby, B. H.; Von Dreele, R. B. GSAS-II: The Genesis of a Modern Open-Source All Purpose Crystallography Software Package. *J Appl Crystallogr* **2013**, *46* (2), 544–549. https://doi.org/10.1107/S0021889813003531.

(34) Stirling, D. R.; Swain-Bowden, M. J.; Lucas, A. M.; Carpenter, A. E.; Cimini, B. A.; Goodman, A. CellProfiler 4: Improvements in Speed, Utility and Usability. *BMC Bioinformatics* **2021**, *22* (1). https://doi.org/10.1186/s12859-021-04344-9.

(35) Wojdyr, M. Fityk: A General-Purpose Peak Fitting Program. *J Appl Crystallogr* **2010**, *43* (5 PART 1), 1126–1128. https://doi.org/10.1107/S0021889810030499.

(36) Mooney, J.; Kambhampati, P. Get the Basics Right: Jacobian Conversion of Wavelength and Energy Scales for Quantitative Analysis of Emission Spectra. *Journal of Physical Chemistry Letters* **2013**, *4* (19), 3316–3318. https://doi.org/10.1021/jz401508t.

(37) Cliffe, M. J.; Goodwin, A. L. PASCal: A Principal Axis Strain Calculator for Thermal Expansion and Compressibility Determination. *J Appl Crystallogr* **2012**, *45* (6), 1321–1329. https://doi.org/10.1107/S0021889812043026.

(38) Van Wyk, L. M.; Loots, L.; Barbour, L. J. Open Access Article. *Chem. Commun* **2021**, *57*, 7693. https://doi.org/10.1039/d1cc01717a.

(39) García-Fernández, A.; Bermúdez-García, J. M.; Castro-García, S.; Llamas-Saiz, A. L.; Artiaga, R.; López-Beceiro, J.; Hu, S.; Ren, W.; Stroppa, A.; Sánchez-Andújar, M.; Señarís-Rodríguez, M. A. Phase Transition, Dielectric Properties, and Ionic Transport in the [(CH3)2NH2]PbI3 Organic-Inorganic Hybrid with 2H-Hexagonal Perovskite Structure. *Inorg Chem* **2017**, *56* (9), 4918–4927. https://doi.org/10.1021/ACS.INORGCHEM.6B03095.

(40) Gervais, F.; Piriou, B. Temperature Dependence of Transverse- and Longitudinal-Optic Modes in TiO2 (Rutile). *Phys Rev B* **1974**, *10* (4), 1642–1654. https://doi.org/10.1103/PhysRevB.10.1642.

(41) Glaser, T.; Müller, C.; Sendner, M.; Krekeler, C.; Semonin, O. E.; Hull, T. D.; Yaffe, O.; Owen, J. S.; Kowalsky, W.; Pucci, A.; Lovrinčić, R. Infrared Spectroscopic Study of Vibrational Modes in Methylammonium Lead Halide Perovskites. *J Phys Chem Lett* **2015**, *6* (15), 2913–2918. https://doi.org/10.1021/acs.jpclett.5b01309.

(42) Boldyrev, K. N.; Anikeeva, V. E.; Semenova, O. I.; Popova, M. N. Infrared Spectra of the CH3NH3PbI3 Hybrid Perovskite: Signatures of Phase Transitions and of Organic Cation Dynamics. *Journal of Physical Chemistry C* **2020**, *124* (42), 23307–23316. https://doi.org/10.1021/acs.jpcc.0c06103.

(43) Tauc, J. Optical Properties and Electronic Structure of Amorphoues Ge and Si. *Mater Res Bull* **1968**, *3* (37–96), 1968.

(44) Mancini, A.; Quadrelli, P.; Amoroso, G.; Milanese, C.; Boiocchi, M.; Sironi, A.; Patrini, M.; Guizzetti, G.; Malavasi, L. Synthesis, Structural and Optical Characterization of APbX3 (A=methylammonium, Dimethylammonium, Trimethylammonium; X=I, Br, Cl) Hybrid Organic-Inorganic Materials. *J Solid State Chem* **2016**, *240*, 55–60. https://doi.org/10.1016/j.jssc.2016.05.015.

(45) Leguy, A. M. A.; Goñi, A. R.; Frost, J. M.; Skelton, J.; Brivio, F.; Rodríguez-Martínez, X.; Weber, O. J.; Pallipurath, A.; Alonso, M. I.; Campoy-Quiles, M.; Weller, M. T.; Nelson, J.; Walsh, A.; Barnes, P. R. F. Dynamic Disorder, Phonon Lifetimes, and the Assignment of Modes to the Vibrational Spectra of Methylammonium Lead Halide Perovskites. *Physical Chemistry Chemical Physics* **2016**, *18* (39), 27051–27066. https://doi.org/10.1039/c6cp03474h.

(46) Mączka, M.; Ptak, M. Temperature-Dependent Raman Studies of FAPbBr3 and MAPbBr3 Perovskites: Effect of Phase Transitions on Molecular Dynamics and Lattice Distortion. *Solids* **2022**, *3* (1), 111–121. https://doi.org/10.3390/solids3010008.

(47) Ibaceta-Jaña, J.; Muydinov, R.; Rosado, P.; Mirhosseini, H.; Chugh, M.; Nazarenko, O.; Dirin, D. N.; Heinrich, D.; Wagner, M. R.; Kühne, T. D.; Szyszka, B.; Kovalenko, M. V.; Hoffmann, A. Vibrational Dynamics in Lead Halide Hybrid Perovskites Investigated by Raman Spectroscopy. *Physical Chemistry Chemical Physics* **2020**, *22* (10), 5604–5614. https://doi.org/10.1039/C9CP06568G.

(48) Nakada, K.; Matsumoto, Y.; Shimoi, Y.; Yamada, K.; Furukawa, Y. Temperature-Dependent Evolution of Raman Spectra of Methylammonium Lead Halide Perovskites, CH3NH3PbX3 (X = I, Br). *Molecules* **2019**, *24* (3). https://doi.org/10.3390/molecules24030626.

(49) Pérez-Osorio, M. A.; Lin, Q.; Phillips, R. T.; Milot, R. L.; Herz, L. M.; Johnston, M. B.; Giustino, F. Raman Spectrum of the Organic-Inorganic Halide Perovskite CH3NH3PbI3 from First Principles and High-Resolution Low-Temperature Raman Measurements. *Journal of Physical Chemistry C* **2018**, *122* (38), 21703–21717. https://doi.org/10.1021/acs.jpcc.8b04669.

(50) Pérez-Osorio, M. A.; Milot, R. L.; Filip, M. R.; Patel, J. B.; Herz, L. M.; Johnston, M. B.; Giustino, F. Vibrational Properties of the Organic–Inorganic Halide Perovskite $CH_3NH_3PbI_3$ from Theory and Experiment: Factor Group Analysis, First-Principles Calculations, and Low-Temperature Infrared Spectra. *The Journal of Physical Chemistry C* **2015**, *119* (46), 25703–25718. https://doi.org/10.1021/acs.jpcc.5b07432.

(51) Niemann, R. G.; Kontos, A. G.; Palles, D.; Kamitsos, E. I.; Kaltzoglou, A.; Brivio, F.; Falaras, P.; Cameron, P. J. Halogen Effects on Ordering and Bonding of $CH_3NH_3^+$ in $CH_3NH_3PbX_3$ (X = Cl, Br, I) Hybrid Perovskites: A Vibrational Spectroscopic Study. *The Journal of Physical Chemistry C* **2016**, *120* (5), 2509–2519. https://doi.org/10.1021/acs.jpcc.5b11256.

(52) Mączka, M.; Zierkiewicz, W.; Michalska, D.; Hanuza, J. Vibrational Properties and DFT Calculations of the Perovskite Metal Formate Framework of [(CH3)2NH2][Ni(HCOO3)] System. *Spectrochim Acta A Mol Biomol Spectrosc* **2014**, *128*, 674–680. https://doi.org/10.1016/j.saa.2014.03.006.

(53) Kontos, A. G.; Manolis, G. K.; Kaltzoglou, A.; Palles, D.; Kamitsos, E. I.; Kanatzidis, M. G.; Falaras, P. Halogen–$NH_2^+$ Interaction, Temperature-Induced Phase Transition, and Ordering in $(NH_2CHNH_2)PbX_3$ (X = Cl, Br,

I) Hybrid Perovskites. *The Journal of Physical Chemistry C* **2020**, *124* (16), 8479–8487. https://doi.org/10.1021/acs.jpcc.9b11334.

(54) García-Fernández, A.; Bermúdez-García, J. M.; Castro-García, S.; Llamas-Saiz, A. L.; Artiaga, R.; López-Beceiro, J.; Hu, S.; Ren, W.; Stroppa, A.; Sánchez-Andújar, M.; Señarís-Rodríguez, M. A. Phase Transition, Dielectric Properties, and Ionic Transport in the [(CH3)2NH2]PbI3 Organic-Inorganic Hybrid with 2H-Hexagonal Perovskite Structure. *Inorg Chem* **2017**, *56* (9), 4918–4927. https://doi.org/10.1021/acs.inorgchem.6b03095.

(55) Fabini, D. H.; Seshadri, R.; Kanatzidis, M. G. The Underappreciated Lone Pair in Halide Perovskites Underpins Their Unusual Properties. *MRS Bull* **2020**, *45* (6), 467–477. https://doi.org/10.1557/mrs.2020.142.

(56) Radha, S. K.; Bhandari, C.; Lambrecht, W. R. L. Distortion Modes in Halide Perovskites: To Twist or to Stretch, a Matter of Tolerance and Lone Pairs. *Phys Rev Mater* **2018**, *2* (6). https://doi.org/10.1103/PhysRevMaterials.2.063605.

(57) García-Fernández, A.; Juarez-Perez, E. J.; Bermúdez-García, J. M.; Llamas-Saiz, A. L.; Artiaga, R.; López-Beceiro, J. J.; Señarís-Rodríguez, M. A.; Sánchez-Andújar, M.; Castro-García, S. Hybrid Lead Halide [(CH 3 ) 2 NH 2 ]PbX 3 (X = Cl − and Br − ) Hexagonal Perovskites with Multiple Functional Properties. *J Mater Chem C Mater* **2019**, *7* (32), 10008–10018. https://doi.org/10.1039/c9tc03543e.

(58) Mączka, M.; Zierkiewicz, W.; Michalska, D.; Hanuza, J. Vibrational Properties and DFT Calculations of the Perovskite Metal Formate Framework of [(CH3)2NH2][Ni(HCOO3)] System. *Spectrochim Acta A Mol Biomol Spectrosc* **2014**, *128*, 674–680. https://doi.org/10.1016/j.saa.2014.03.006.

(59) Svane, K. L.; Forse, A. C.; Grey, C. P.; Kieslich, G.; Cheetham, A. K.; Walsh, A.; Butler, K. T. How Strong Is the Hydrogen Bond in Hybrid Perovskites? *Journal of Physical Chemistry Letters* **2017**, *8* (24), 6154–6159. https://doi.org/10.1021/acs.jpclett.7b03106.

(60) Grechko, M.; Bretschneider, S. A.; Vietze, L.; Kim, H.; Bonn, M. Vibrational Coupling between Organic and Inorganic Sublattices of Hybrid Perovskites. *Angewandte Chemie - International Edition* **2018**, *57* (41), 13657–13661. https://doi.org/10.1002/anie.201806676.

(61) El-Mellouhi, F.; Marzouk, A.; Bentria, E. T.; Rashkeev, S. N.; Kais, S.; Alharbi, F. H. Hydrogen Bonding and Stability of Hybrid Organic–Inorganic Perovskites. *ChemSusChem* **2016**, *9* (18), 2648–2655. https://doi.org/10.1002/cssc.201600864.

(62) Kieslich, G.; Skelton, J. M.; Armstrong, J.; Wu, Y.; Wei, F.; Svane, K. L.; Walsh, A.; Butler, K. T. Hydrogen Bonding versus Entropy: Revealing the Underlying Thermodynamics of the Hybrid Organic-Inorganic Perovskite [CH3NH3]PbBr3. *Chemistry of Materials* **2018**, *30* (24), 8782–8788. https://doi.org/10.1021/acs.chemmater.8b03164.

(63) El-Mellouhi, F.; Marzouk, A.; Bentria, E. T.; Rashkeev, S. N.; Kais, S.; Alharbi, F. H. Hydrogen Bonding and Stability of Hybrid Organic–Inorganic Perovskites. *ChemSusChem* **2016**, *9* (18), 2648–2655. https://doi.org/10.1002/cssc.201600864.

(64) Tian, C.; Liang, Y.; Chen, W.; Huang, Y.; Huang, X.; Tian, F.; Yang, X. Hydrogen-Bond Enhancement Triggered Structural Evolution and Band Gap Engineering of Hybrid Perovskite (C6H5CH2NH3)2PbI4 under High Pressure. *Physical Chemistry Chemical Physics* **2020**, *22* (4), 1841–1846. https://doi.org/10.1039/c9cp05904k.

(65) Svane, K. L.; Forse, A. C.; Grey, C. P.; Kieslich, G.; Cheetham, A. K.; Walsh, A.; Butler, K. T. How Strong Is the Hydrogen Bond in Hybrid Perovskites? *Journal of Physical Chemistry Letters* **2017**, *8* (24), 6154–6159. https://doi.org/10.1021/acs.jpclett.7b03106.

(66) Carabatos-Nédelec, C.; Becker, P. Order-Disorder and Structural Phase Transitions in Solid-State Materials by Raman Scattering Analysis. *Journal of Raman Spectroscopy* **1997**, *28* (9), 663–671. https://doi.org/10.1002/(sici)1097-4555(199709)28:9<663::aid-jrs157>3.0.co;2-l.

(67) Mączka, M.; Ptak, M.; Vasconcelos, D. L. M.; Giriunas, L.; Freire, P. T. C.; Bertmer, M.; Banys, J.; Simenas, M. NMR and Raman Scattering Studies of Temperature- and Pressure-Driven Phase Transitions in $CH_3NH_2NH_2PbCl_3$ Perovskite. *The Journal of Physical Chemistry C* **2020**, *124* (49), 26999–27008. https://doi.org/10.1021/acs.jpcc.0c07886.

(68) García-Fernández, A.; Bermúdez-García, J. M.; Castro-García, S.; Llamas-Saiz, A. L.; Artiaga, R.; López-Beceiro, J. J.; Sánchez-Andújar, M.; Señarís-Rodríguez, M. A. $[(CH_3)_2NH_2]_7Pb_4X_{15}$ ($X = Cl^-$ and $Br^-$), 2D-Perovskite Related Hybrids with Dielectric Transitions and Broadband Photoluminiscent Emission. *Inorg Chem* **2018**, *57* (6), 3215–3222. https://doi.org/10.1021/acs.inorgchem.7b03217.

(69) Zafar, Z.; Zafar, A.; Guo, X.; Lin, Q.; Yu, Y. Raman Evolution of Order–Disorder Phase Transition in Multiaxial Molecular Ferroelectric Thin Film. *Journal of Raman Spectroscopy* **2019**, *50* (10), 1576–1583. https://doi.org/10.1002/jrs.5642.

(70) Mączka, M.; Gągor, A.; Macalik, B.; Pikul, A.; Ptak, M.; Hanuza, J. Order–Disorder Transition and Weak Ferromagnetism in the Perovskite Metal Formate Frameworks of [(CH3)2 NH2][M(HCOO)3] and [(CH3) 2 ND2

][M(HCOO)3] (M = Ni, Mn). *Inorg Chem* **2014**, *53* (1), 457–467. https://doi.org/10.1021/ic402425n.

(71) Samet, A.; Ahmed, A. Ben; Mlayah, A.; Boughzala, H.; Hlil, E. K.; Abid, Y. Optical Properties and Ab Initio Study on the Hybrid Organic-Inorganic Material [(CH3)2NH2]3[BiI 6]. *J Mol Struct* **2010**, *977* (1–3), 72–77. https://doi.org/10.1016/j.molstruc.2010.05.016.

(72) Rok, M.; Bator, G.; Zarychta, B.; Dziuk, B.; Repeć, J.; Medycki, W.; Zamponi, M.; Usevičius, G.; Šimenas, M.; Banys, J. Isostructural Phase Transition, Quasielastic Neutron Scattering and Magnetic Resonance Studies of a Bistable Dielectric Ion-Pair Crystal [(CH 3 ) 2 NH 2 ] 2 KCr(CN) 6. *Dalton Transactions* **2019**, *48* (13), 4190–4202. https://doi.org/10.1039/c8dt05082a.

(73) Ri, ardas Sobiestianskas, Kohji Abe, and T. S. Line Shape of the Raman Spectrum and the Disorder Mechanism of the Methylammonium Group in a [(CH 3) 2NH 2] 5Cd 3Cl 11 Crystal. *J Physical Soc Japan* **1996**, *65* (10), 3146–3149. https://doi.org/https://doi.org/10.1143/JPSJ.65.3146.

(74) Asaji, T.; Ashitomi, K. Phase Transition and Cationic Motion in a Metal-Organic Perovskite, Dimethylammonium Zinc Formate [(CH3)2NH 2][Zn(HCOO)3]. *Journal of Physical Chemistry C* **2013**, *117* (19), 10185–10190. https://doi.org/10.1021/jp402148y.

(75) Mączka, M.; Pietraszko, A.; MacAlik, L.; Sieradzki, A.; Trzmiel, J.; Pikul, A. Synthesis and Order-Disorder Transition in a Novel Metal Formate Framework of [(CH<inf>3</Inf>)<inf>2</Inf>NH<inf>2</Inf>]Na<inf>0.5</Inf>Fe<inf>0.5</Inf>(HCOO)<inf>3</Inf>]. *Dalton Transactions* **2014**, *43* (45). https://doi.org/10.1039/c4dt02586e.

(76) Trzebiatowska, M.; Maczka, M.; Ptak, M.; Giriunas, L.; Balciunas, S.; Simenas, M.; Klose, D.; Banys, J. Spectroscopic Study of Structural Phase Transition and Dynamic Effects in a [(CH3)2NH2][Cd(N3)3] Hybrid Perovskite Framework. *Journal of Physical Chemistry C* **2019**, *123* (18), 11840–11849. https://doi.org/10.1021/ACS.JPCC.9B01121/SUPPL_FILE/JP9B01121_SI_001.PDF.

(77) Du, Z. Y.; Sun, Y. Z.; Chen, S. L.; Huang, B.; Su, Y. J.; Xu, T. T.; Zhang, W. X.; Chen, X. M. Insight into the Molecular Dynamics of Guest Cations Confined in Deformable Azido Coordination Frameworks. *Chemical Communications* **2015**, *51* (86), 15641–15644. https://doi.org/10.1039/C5CC06863K.

(78) Stoumpos, C. C.; Malliakas, C. D.; Kanatzidis, M. G. Semiconducting Tin and Lead Iodide Perovskites with Organic Cations: Phase Transitions, High

Mobilities, and near-Infrared Photoluminescent Properties. *Inorg Chem* **2013**, *52* (15), 9019–9038. https://doi.org/10.1021/ic401215x.

(79) Franssen, W. M. J.; Bruijnaers, B. J.; Portengen, V. H. L.; Kentgens, A. P. M. Dimethylammonium Incorporation in Lead Acetate Based MAPbI3 Perovskite Solar Cells. *ChemPhysChem* **2018**, *19* (22), 3107–3115. https://doi.org/10.1002/cphc.201800732.

(80) Kim, B.; Kim, M.; Kim, H.; Jeong, S.; Yang, J.; Jeong, M. S. Improved Stability of MAPbI$_3$ Perovskite Solar Cells Using Two-Dimensional Transition-Metal Dichalcogenide Interlayers. *ACS Appl Mater Interfaces* **2022**, *14* (31), 35726–35733. https://doi.org/10.1021/acsami.2c08680.

(81) Li, J.; Wang, H.; Li, D. Self-Trapped Excitons in Two-Dimensional Perovskites. *Frontiers of Optoelectronics* **2020**, *13* (3), 225–234. https://doi.org/10.1007/S12200-020-1051-X/METRICS.

(82) Mamaeva, M. P.; Androulidaki, M.; Spanou, V.; Kireev, N. M.; Pelekanos, N. T.; Kapitonov, Y. V.; Stoumpos, C. Free Exciton and Defect-Related States in CH3NH3PbCl3 Perovskite Single Crystal. *Journal of Physical Chemistry C* **2023**, *127* (46), 22784–22789. https://doi.org/10.1021/ACS.JPCC.3C05829/SUPPL_FILE/JP3C05829_SI_001.PDF.

(83) Basumatary, P.; Kumari, J.; Agarwal, P. Probing the Defects States in MAPbI3 Perovskite Thin Films through Photoluminescence and Photoluminescence Excitation Spectroscopy Studies. *Optik (Stuttg)* **2022**, *266*, 169586. https://doi.org/10.1016/J.IJLEO.2022.169586.

(84) Song, K. S.; Williams, R. T. Self-Trapped Excitons. **1996**, *105*. https://doi.org/10.1007/978-3-642-85236-7.

(85) Crace, E. J.; Su, A. C.; Karunadasa, H. I. Reliably Obtaining White Light from Layered Halide Perovskites at Room Temperature. *Chem Sci* **2022**, *13* (34), 9973–9979. https://doi.org/10.1039/D2SC02381D.

(86) Quan, L. N.; Yang, P.; Guo, P. Octahedral Distortion and Excitonic Behavior of Cs3Bi2Br9 Halide Perovskite at Low Temperature. *Journal of Physical Chemistry C* **2022**. https://doi.org/10.1021/ACS.JPCC.2C07642/SUPPL_FILE/JP2C07642_SI_001.PDF.

(87) Wright, A. D.; Verdi, C.; Milot, R. L.; Eperon, G. E.; Pérez-Osorio, M. A.; Snaith, H. J.; Giustino, F.; Johnston, M. B.; Herz, L. M. Electron–Phonon Coupling in Hybrid Lead Halide Perovskites. *Nature Communications 2016 7:1* **2016**, *7* (1), 1–9. https://doi.org/10.1038/ncomms11755.

(88) Yamada, Y.; Kanemitsu, Y. Electron-Phonon Interactions in Halide Perovskites. *NPG Asia Materials 2022 14:1* **2022**, *14* (1), 1–15. https://doi.org/10.1038/s41427-022-00394-4.

(89) Tan, J.; Li, D.; Zhu, J.; Han, N.; Gong, Y.; Zhang, Y. Self-Trapped Excitons in Soft Semiconductors. *Nanoscale* **2022**, *14* (44), 16394–16414. https://doi.org/10.1039/D2NR03935D.

(90) Pan, F.; Li, J.; Ma, X.; Nie, Y.; Liu, B.; Ye, H. Free and Self-Trapped Exciton Emission in Perovskite $CsPbBr_3$ Microcrystals. *RSC Adv* **2022**, *12* (2), 1035–1042. https://doi.org/10.1039/D1RA08629D.

(91) McCall, K. M.; Stoumpos, C. C.; Kostina, S. S.; Kanatzidis, M. G.; Wessels, B. W. Strong Electron-Phonon Coupling and Self-Trapped Excitons in the Defect Halide Perovskites $A_3M_2I_9$ (A = Cs, Rb; M = Bi, Sb). *Chemistry of Materials* **2017**, *29* (9), 4129–4145. https://doi.org/10.1021/acs.chemmater.7b01184.

(92) Xu, H.; Zhang, Z.; Dong, X.; Huang, L.; Zeng, H.; Lin, Z.; Zou, G. Corrugated 1D Hybrid Metal Halide $[C_6H_7ClN]CdCl_3$ Exhibiting Broadband White-Light Emission. *Inorg Chem* **2022**, *61* (11), 4752–4759. https://doi.org/10.1021/acs.inorgchem.2c00169.

(93) Klement, P.; Dehnhardt, N.; Dong, C. D.; Dobener, F.; Bayliff, S.; Winkler, J.; Hofmann, D. M.; Klar, P. J.; Schumacher, S.; Chatterjee, S.; Heine, J. Atomically Thin Sheets of Lead-Free 1D Hybrid Perovskites Feature Tunable White-Light Emission from Self-Trapped Excitons. *Advanced Materials* **2021**, *33* (23). https://doi.org/10.1002/adma.202100518.

(94) Liu, S.; Du, Y. W.; Tso, C. Y.; Lee, H. H.; Cheng, R.; Feng, S. P.; Yu, K. M. Organic Hybrid Perovskite ($MAPbI_{3-x}Cl_x$) for Thermochromic Smart Window with Strong Optical Regulation Ability, Low Transition Temperature, and Narrow Hysteresis Width. *Adv Funct Mater* **2021**, *31* (26). https://doi.org/10.1002/adfm.202010426.

(95) Roy, A.; Ullah, H.; Ghosh, A.; Baig, H.; Sundaram, S.; Tahir, A. A.; Mallick, T. K. Understanding the Semi-Switchable Thermochromic Behavior of Mixed Halide Hybrid Perovskite Nanorods. *Journal of Physical Chemistry C* **2021**, *125* (32), 18058–18070. https://doi.org/10.1021/acs.jpcc.1c05487.

(96) Han, J. H.; Samanta, T.; Park, Y. M.; Cho, H. Bin; Min, J. W.; Hwang, S. J.; Jang, S. W.; Im, W. Bin. Effect of Self-Trapped Excitons in the Optical Properties of Manganese-Alloyed Hexagonal-Phased Metal Halide Perovskite. *Chemical Engineering Journal* **2022**, *450*. https://doi.org/10.1016/j.cej.2022.138325.

(97) Xu, H.; Zhang, Z.; Dong, X.; Huang, L.; Zeng, H.; Lin, Z.; Zou, G. Corrugated 1D Hybrid Metal Halide [C6H7ClN]CdCl3 Exhibiting Broadband White-Light Emission. *Inorg Chem* **2022**, *61* (11), 4752–4759. https://doi.org/10.1021/ACS.INORGCHEM.2C00169/SUPPL_FILE/IC2C00169_SI_001.PDF.

(98) McCall, K. M.; Stoumpos, C. C.; Kostina, S. S.; Kanatzidis, M. G.; Wessels, B. W. Strong Electron-Phonon Coupling and Self-Trapped Excitons in the Defect Halide Perovskites A3M2I9 (A = Cs, Rb; M = Bi, Sb). *Chemistry of Materials* **2017**, *29* (9), 4129–4145. https://doi.org/10.1021/ACS.CHEMMATER.7B01184/SUPPL_FILE/CM7B01184_SI_002.CIF.

(99) Pan, F.; Li, J.; Ma, X.; Nie, Y.; Liu, B.; Ye, H. Free and Self-Trapped Exciton Emission in Perovskite CsPbBr3 Microcrystals. *RSC Adv* **2021**, *12* (2), 1035–1042. https://doi.org/10.1039/D1RA08629D.

(100) Han, Y.; Cheng, X.; Cui, B. Bin. Factors Influencing Self-Trapped Exciton Emission of Low-Dimensional Metal Halides. *Mater Adv* **2023**, *4* (2), 355–373. https://doi.org/10.1039/D2MA00676F.

(101) Tong, Y. B.; Ren, L. Te; Duan, H. B.; Liu, J. L.; Ren, X. M. An Amphidynamic Inorganic–Organic Hybrid Crystal of Bromoplumbate with 1,5-Bis(1-Methylimidazolium)Pentane Exhibiting Multi-Functionality of a Dielectric Anomaly and Temperature-Dependent Dual Band Emissions. *Dalton Transactions* **2015**, *44* (40), 17850–17858. https://doi.org/10.1039/C5DT02739J.

(102) Li, Y. Y.; Lin, C. K.; Zheng, G. L.; Cheng, Z. Y.; You, H.; Wang, W. D.; Lin, J. Novel ⟨110⟩-Oriented Organic-Inorganic Perovskite Compound Stabilized by N-(3-Aminopropyl)Imidazole with Improved Optical Properties. *Chemistry of Materials* **2006**, *18* (15), 3463–3469. https://doi.org/10.1021/CM060714U/SUPPL_FILE/CM060714USI20060429_054859.PDF.

(103) Nowick, A. S.; Du, Y.; Liang, K. C. Some Factors That Determine Proton Conductivity in Nonstoichiometric Complex Perovskites. *Solid State Ion* **1999**, *125* (1–4), 303–311. https://doi.org/10.1016/S0167-2738(99)00189-7.

(104) Mao, L.; Guo, P.; Kepenekian, M.; Hadar, I.; Katan, C.; Even, J.; Schaller, R. D.; Stoumpos, C. C.; Kanatzidis, M. G. Structural Diversity in White-Light-Emitting Hybrid Lead Bromide Perovskites. *J Am Chem Soc* **2018**, *140* (40), 13078–13088. https://doi.org/10.1021/JACS.8B08691/SUPPL_FILE/JA8B08691_SI_008.CIF.

(105) Smith, M. D.; Karunadasa, H. I. White-Light Emission from Layered Halide Perovskites. *Acc Chem Res* **2018**, *51* (3), 619–627. https://doi.org/10.1021/acs.accounts.7b00433.

(106) Zhao, J. Q.; Jing, C. Q.; Wu, J. H.; Zhang, W. F.; Feng, L. J.; Yue, C. Y.; Lei, X. W. Systematic Approach of One-Dimensional Lead Perovskites with Face-Sharing Connectivity to Realize Efficient and Tunable Broadband Light Emission. *Journal of Physical Chemistry C* **2021**, *125* (20), 10850–10859. https://doi.org/10.1021/acs.jpcc.1c00515.

(107) Yuan, Z.; Zhou, C.; Tian, Y.; Shu, Y.; Messier, J.; Wang, J. C.; Van De Burgt, L. J.; Kountouriotis, K.; Xin, Y.; Holt, E.; Schanze, K.; Clark, R.; Siegrist, T.; Ma, B. One-Dimensional Organic Lead Halide Perovskites with Efficient Bluish White-Light Emission. *Nat Commun* **2017**, *8*. https://doi.org/10.1038/NCOMMS14051.

(108) Franchini, C.; Reticcioli, M.; Setvin, M.; Diebold, U. Polarons in Materials. *Nature Reviews Materials 2021 6:7* **2021**, *6* (7), 560–586. https://doi.org/10.1038/s41578-021-00289-w.

(109) Yin, J.; Li, H.; Cortecchia, D.; Soci, C.; Brédas, J. L. Excitonic and Polaronic Properties of 2D Hybrid Organic-Inorganic Perovskites. *ACS Energy Lett* **2017**, *2* (2), 417–423. https://doi.org/10.1021/ACSENERGYLETT.6B00659.

(110) Zhu, P.; Zhu, H.; Adhikari, G. C.; Thapa, S. Spectral Optimization of White Light from Hybrid Metal Halide Perovskites. *OSA Contin* **2019**, *2* (6), 1880. https://doi.org/10.1364/osac.2.001880.